\documentclass[prl,amsfonts,twoside,amssymb,superscriptaddress,twocolumn]{revtex4-2}
\usepackage[normalem]{ulem}
\usepackage[utf8]{inputenc}
\usepackage[T1]{fontenc}
\usepackage{ragged2e}
\usepackage{dcolumn}
\usepackage{amsmath}
\usepackage{epsfig}
\usepackage{float}
\usepackage{graphicx}
\usepackage{subfigure}
\usepackage{hyperref}
\usepackage{colortbl}
\usepackage{multirow}
\usepackage{amsthm,amsmath,amssymb}
\usepackage{mathrsfs}
\usepackage{bm}
\usepackage{enumitem}
\definecolor{Forest}{RGB}{34,139,34}
\graphicspath{{./images/}}

\begin{document}

\title{A polychromatic continuous-variable quantum communication network enabled by optical frequency combs}

\author{Yuehan Xu}
\affiliation{State Key Laboratory of Photonics and Communications, Center of Quantum Sensing and Information Processing, Shanghai Jiao Tong University, Shanghai 200240, China}
\author{Qijun Zhang}
\affiliation{State Key Laboratory of Photonics and Communications, Center of Quantum Sensing and Information Processing, Shanghai Jiao Tong University, Shanghai 200240, China}
\author{Junpeng Zhang}
\affiliation{State Key Laboratory of Photonics and Communications, Center of Quantum Sensing and Information Processing, Shanghai Jiao Tong University, Shanghai 200240, China}
\author{Xiaojuan Liao}
\affiliation{State Key Laboratory of Photonics and Communications, Center of Quantum Sensing and Information Processing, Shanghai Jiao Tong University, Shanghai 200240, China}
\author{Ziyi Shen}
\affiliation{State Key Laboratory of Photonics and Communications, Center of Quantum Sensing and Information Processing, Shanghai Jiao Tong University, Shanghai 200240, China}
\author{Xu Liu}
\affiliation{State Key Laboratory of Photonics and Communications, Center of Quantum Sensing and Information Processing, Shanghai Jiao Tong University, Shanghai 200240, China}
\author{Beibei Zhang}
\affiliation{State Key Laboratory of Photonics and Communications, Center of Quantum Sensing and Information Processing, Shanghai Jiao Tong University, Shanghai 200240, China}
\author{Zicong Tan}
\affiliation{State Key Laboratory of Photonics and Communications, Center of Quantum Sensing and Information Processing, Shanghai Jiao Tong University, Shanghai 200240, China}
\author{Zehao Zhou}
\affiliation{State Key Laboratory of Photonics and Communications, Center of Quantum Sensing and Information Processing, Shanghai Jiao Tong University, Shanghai 200240, China}
\author{Jisheng Dai}
\affiliation{College of Information Science and Technology, Donghua University, Shanghai 201620, China}
\author{Xueqin Jiang}
\affiliation{College of Information Science and Technology, Donghua University, Shanghai 201620, China}
\affiliation{Hefei National Laboratory, Hefei 230088, China}
\author{Peng Huang}
\affiliation{State Key Laboratory of Photonics and Communications, Center of Quantum Sensing and Information Processing, Shanghai Jiao Tong University, Shanghai 200240, China}
\affiliation{Hefei National Laboratory, Hefei 230088, China}
\affiliation{Shanghai Research Center for Quantum Sciences, Shanghai 201315, China}
\author{Tao Wang}
\email{tonystar@sjtu.edu.cn}
\affiliation{State Key Laboratory of Photonics and Communications, Center of Quantum Sensing and Information Processing, Shanghai Jiao Tong University, Shanghai 200240, China}
\affiliation{Hefei National Laboratory, Hefei 230088, China}
\affiliation{Shanghai Research Center for Quantum Sciences, Shanghai 201315, China}
\author{Guihua Zeng}
\email{ghzeng@sjtu.edu.cn}
\affiliation{State Key Laboratory of Photonics and Communications, Center of Quantum Sensing and Information Processing, Shanghai Jiao Tong University, Shanghai 200240, China}
\affiliation{Hefei National Laboratory, Hefei 230088, China}
\affiliation{Shanghai Research Center for Quantum Sciences, Shanghai 201315, China}

\begin{abstract}
In classical communication, the introduction of polychromatic resources has rapidly boosted classical networks' rate and scale. Quantum communication is now at a similar critical stage in its development, and therefore, it is essential to investigate polychromatic quantum communication networks. In this letter, we report a polychromatic continuous-variable quantum communication network enabled by optical frequency combs. The multi-mode density matrices constituted by polychromatic quantum networks are studied. Considering the limited mode isolation, the maximum amount of information that eavesdroppers can obtain is recalculated, therefore, the total secret key rate is provided. We have also demonstrated that, compared to other multiplexing techniques, polychromatic quantum networks can theoretically achieve a secret key rate without decreasing with the increase in users. In the experiment, direct-transmission type and round-trip type quantum communication networks were built using optical frequency combs and dual-comb interference detection technology. The Gaussian-modulated continuous-variable quantum key distribution (CV-QKD) protocol has been validated, with a network capacity of $19$ and a total secret key rate of $8.75$ $\rm{Gbps}$ at a uniform distance of $5$ $\rm{km}$ (asymptotic case), $0.82$ $\rm{Mbps}$ at $120$ $\rm{km}$ (finite-size effect), $89.10$ $\rm{Mbps}$ at $40$ $\rm{km}$ (compsable security), $13.66$ $\rm{Mbps}$ at $40$ $\rm{km}$ (compsable finite-size security). This implementation not only provides technical support for a high-speed multi-node quantum network, but also provides a solution for the future quantum Internet with continuous variables.
\end{abstract}

\maketitle

\section{Introduction}

Quantum networks, simply put, are networks that transmit quantum states. Transmitting different quantum states, using different protocols, and applying them in different scenarios all lead to varying but groundbreaking developments. It has already been demonstrated by numerous outstanding works, such as quantum communication networks achieving theoretically secure encryption \cite{townsend1997quantum,duan2001long,choi2011quantum,frohlich2013quantum,frohlich2015quantum,liao2017satellite,wengerowsky2018entanglement,dynes2019cambridge,joshi2020trusted,chen2021integrated,qi202115,chen2021implementation,huang2021realizing,wang2021practical,wang2023experimental}, quantum sensing networks bringing about more precise measurements \cite{proctor2018multiparameter,eldredge2018optimal,xia2020demonstration,liu2021distributed}, and quantum computing networks constructing high-performance computations \cite{cirac1999distributed,jiang2007distributed,van2010distributed,monroe2014large}.

In quantum communication networks, secure quantum key distribution (QKD) can be achieved because the maximum amount of information that eavesdroppers can obtain can be quantitatively estimated and deducted \cite{bennett2014quantum,lucamarini2018overcoming,yin2020entanglement,wang2022twin,fan2022robust,liu2023experimental,li2023high,huang2024cost}. Basic transmission units in a quantum communication network are quantum bits (qubits) or quantum modes (qumodes). They can be physically represented in discrete-variable or continuous-variable quantum systems. In continuous-variable quantum systems, the information is encoded in optical quadrature components, on which any eavesdropping behavior will cause changes. So far, long-distance, high-speed, integrated, and networked continuous-variable QKD (CV-QKD) based on monochromatic resources has been achieved \cite{grosshans2002continuous,qi2015generating,huang2016long,zhang2020long,wang2022sub,xu2023round,hajomer2024continuous,liu2024integrated,hajomer2024long,research.0416,qi2024experimental,pan2024high,hajomer2024continuous,li2024experimental}.

\begin{figure*}
\centering
\subfigure[]{\label{PM}\includegraphics[width=1\linewidth]{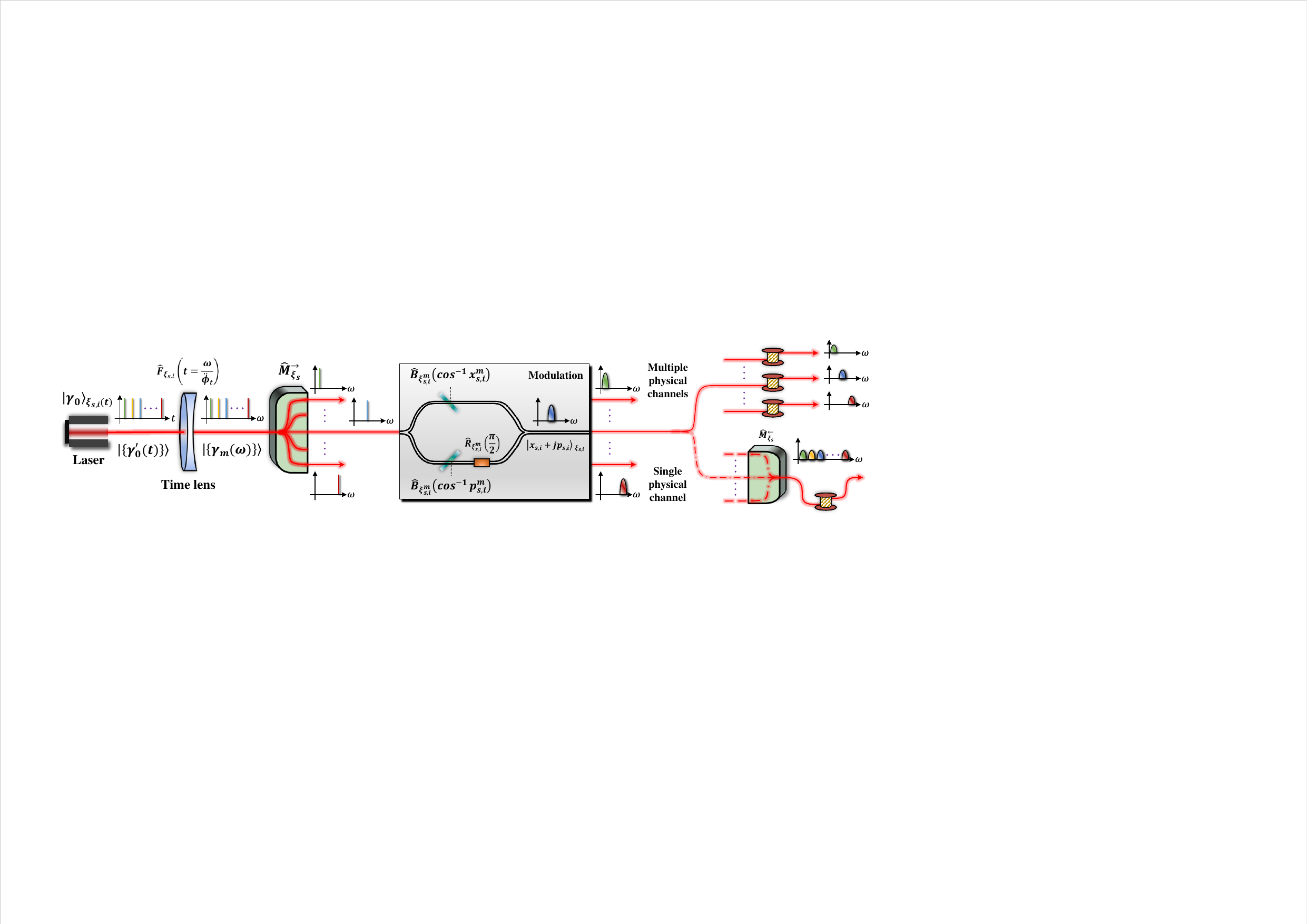}}
\subfigure[]{\label{Topology}\includegraphics[width=0.29\linewidth]{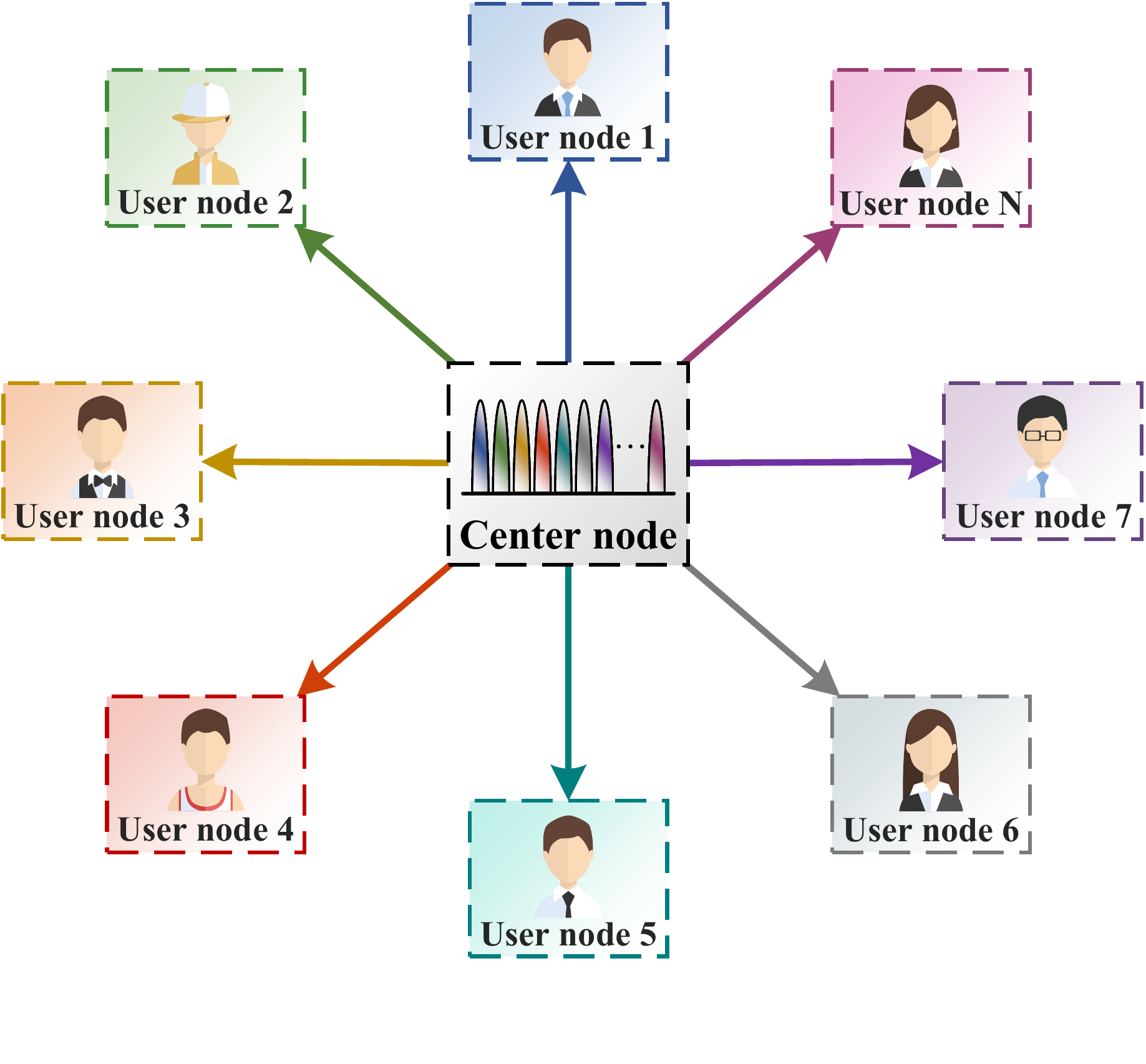}}
\subfigure[]{\label{EB}\includegraphics[width=0.7\linewidth]{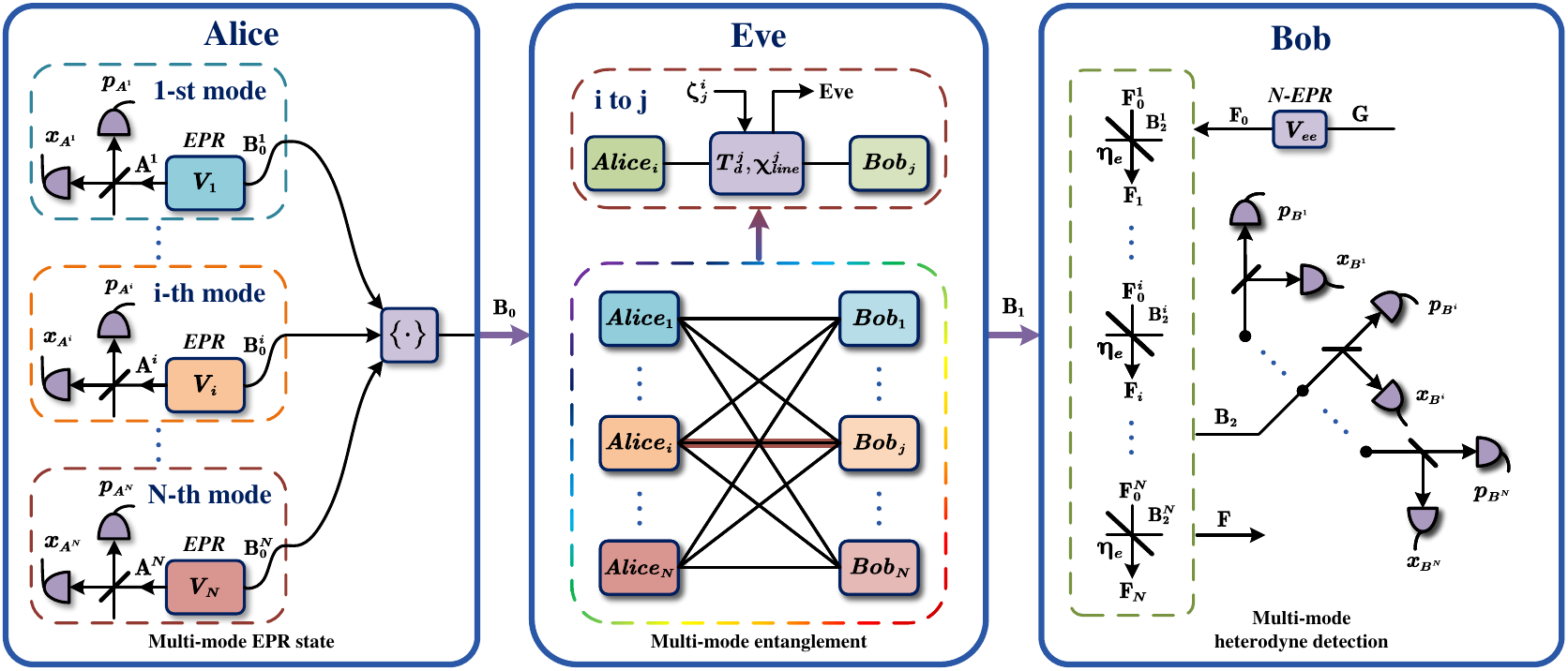}}
\caption{The PM model, topology, and EB model of polychromatic quantum communication network. (a) Multi-mode PM model. (b) Star-topology quantum communication network. (c) Multi-mode EB model.}
\label{PM-EB}
\end{figure*}

Since light offers various degrees of freedom, for example, path, time, frequency, and spatial modes, it presents a promising approach to creating large-scale quantum communication networks. In the closely related field of optical communication, classical optical networks have witnessed substantial enhancements in transmission rate and coverage scope, owing to the transition from monochromatic to polychromatic carriers \cite{brackett1990dense,banerjee2005wavelength,winzer2015scaling}. Consequently, it is imperative to prioritize the evolution from monochromatic to polychromatic quantum networks, which is a crucial step toward scaling it for widespread applications \cite{menicucci2008one,foltynowicz2011quantum,roslund2014wavelength,pfister2019continuous,cai2021quantum,caldwell2022time,liu2024creation,wang2025large,jia2025continuous,fan2025quantum}. At present, a polychromatic continuous-variable quantum communication network with optical frequency combs has not been reported, to our knowledge.

\begin{figure*}
	\centering
	\subfigure[]{\label{SKR-se-NC}\includegraphics[width=0.32\linewidth]{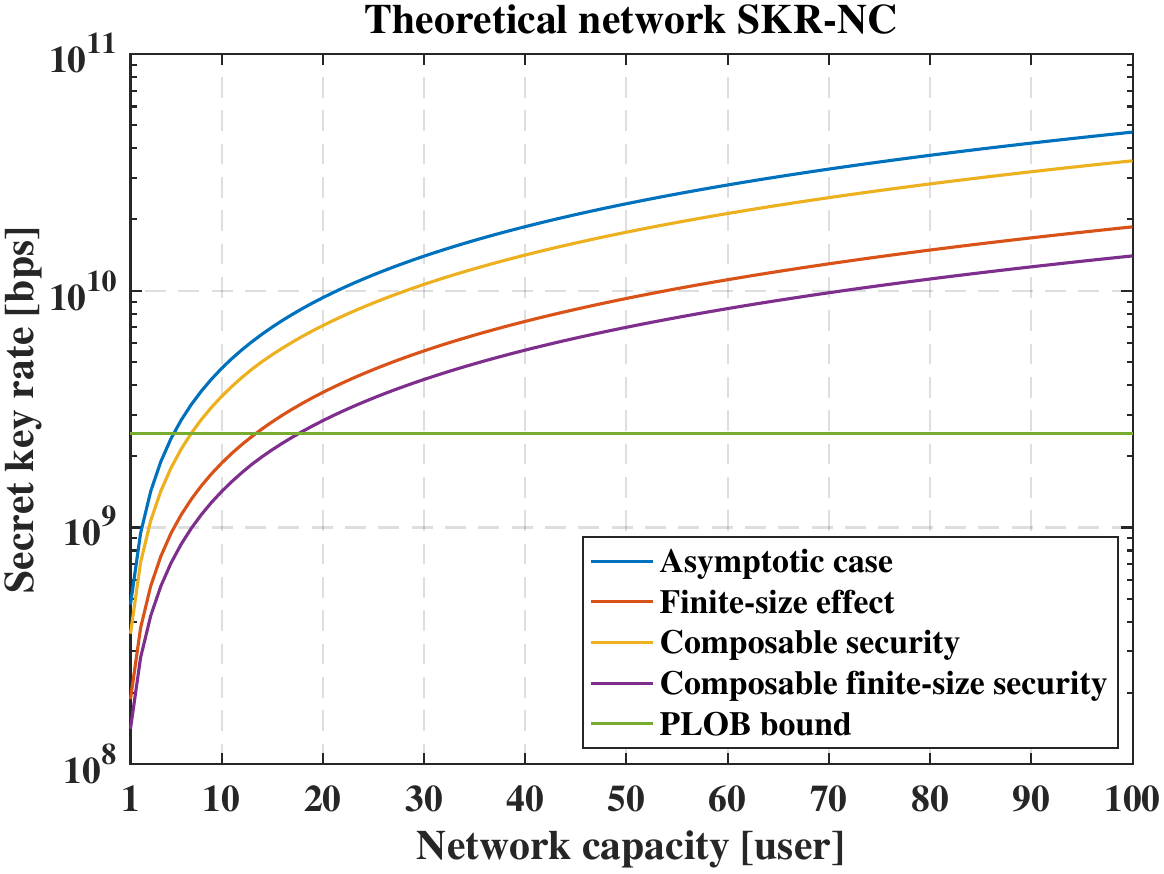}}
	\subfigure[]{\label{SKR-se-TD}\includegraphics[width=0.32\linewidth]{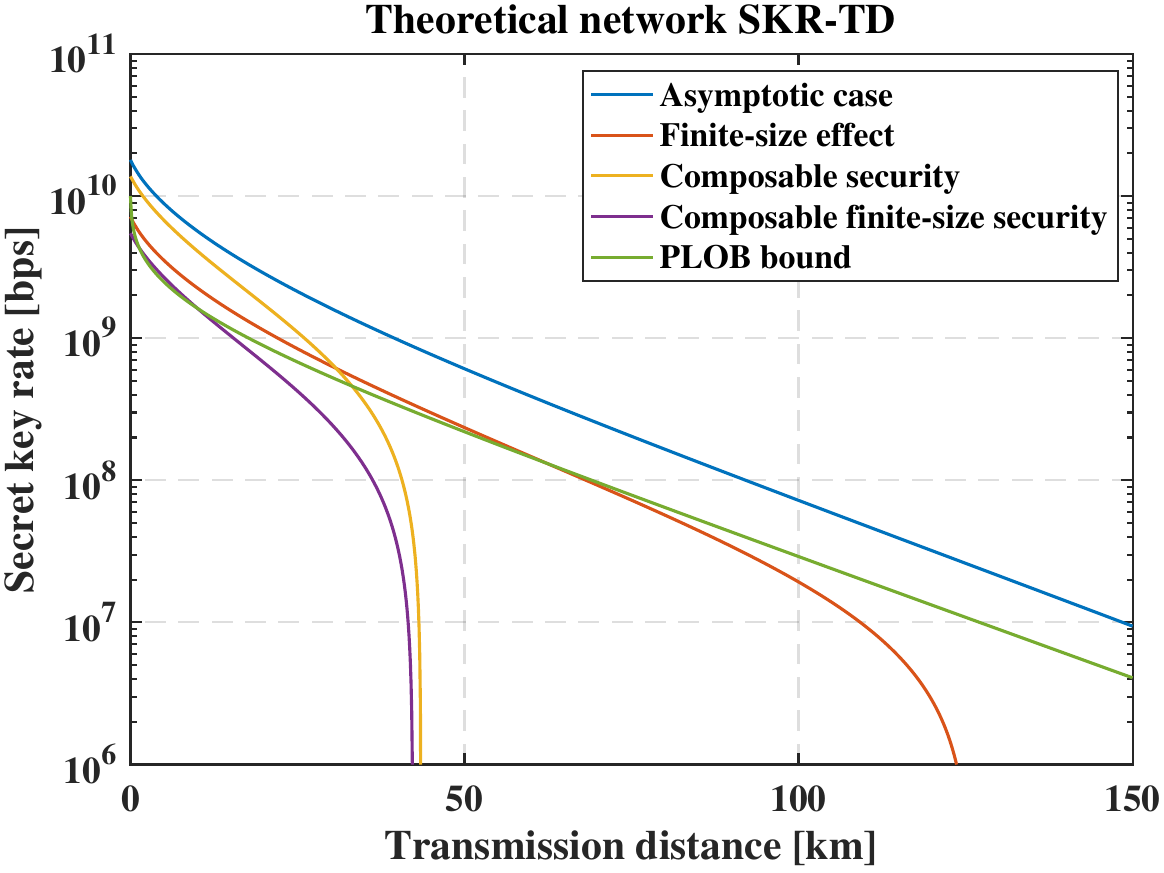}}
	\subfigure[]{\label{SKR-se-MI}\includegraphics[width=0.32\linewidth]{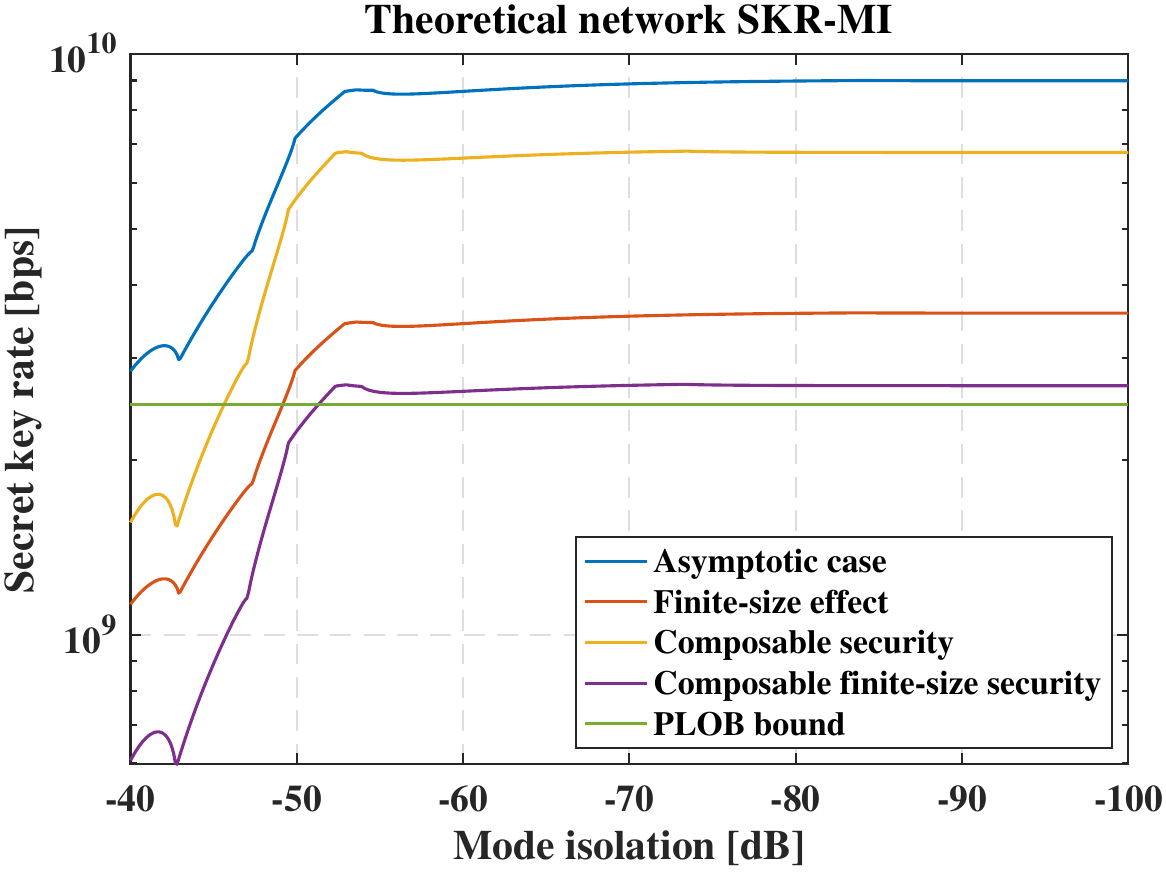}}
	\subfigure[]{\label{SKR-ty-NC}\includegraphics[width=0.49\linewidth]{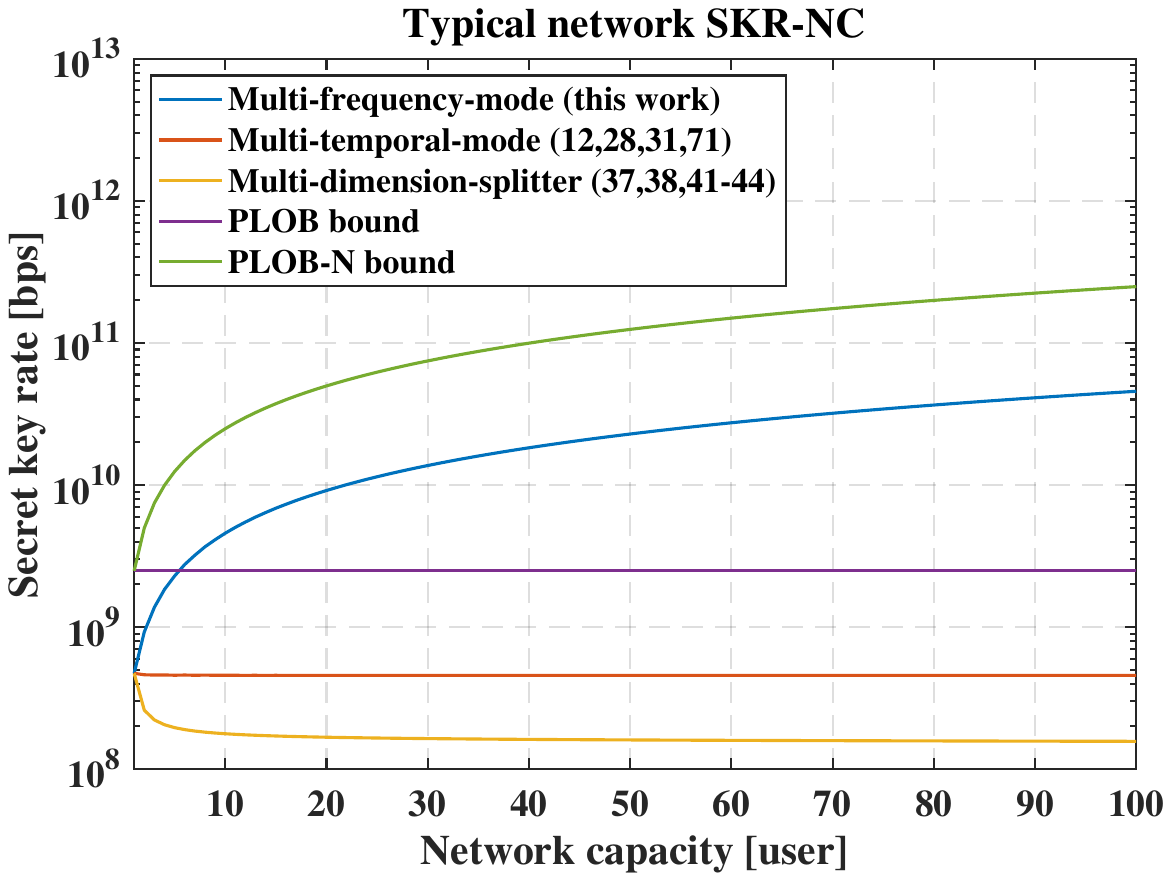}}
	\subfigure[]{\label{SKR-ty-TD}\includegraphics[width=0.49\linewidth]{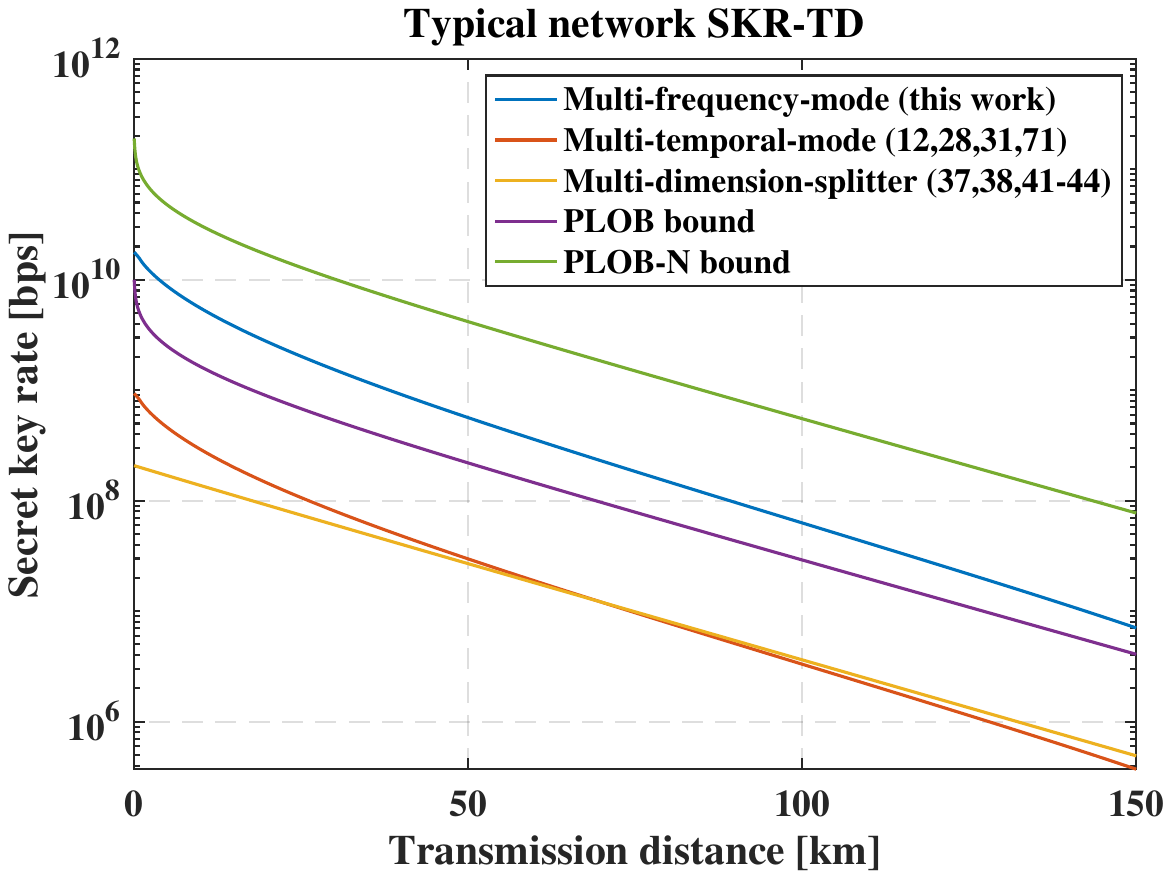}}
	\caption{The relationships between network SKR and various parameters. (a) Theoretical network SKR with network capacity. (b) Theoretical network SKR with transmission distance. (c) Theoretical network SKR with mode isolation. (d) The relationship between SKR and network capacity in different multi-mode quantum networks. (e) The relationship between SKR and transmission distance in different multi-mode quantum networks.}
	\label{SKR-se-ty}
\end{figure*}

In this letter, we report a polychromatic continuous-variable quantum communication network enabled by optical frequency combs. First, the prepare-and-measure (PM) model of a polychromatic quantum network was established based on continuous-mode quantum optics. Second, the corresponding entanglement-based (EB) model was proposed. In terms of information security, it has been observed that the information Eve can obtain in a QKD network is not equivalent to a linear combination of point-to-point link information; rather, it depends on mode isolation. We have also demonstrated that, compared to other multiplexing techniques, polychromatic quantum networks can theoretically achieve a secret key rate (SKR) without decreasing with the increase of users. Last, two types of experimental verifications, direct-transmission type quantum network with local local oscillator (DT-LLO) and round-trip type quantum network with transmitted local oscillator (RT-TLO), are conducted, and both Gaussian-modulated optical frequency comb and dual-comb interference detection technology have been developed. The experimental results demonstrate that the network can achieve a network capacity of $19$, and the achievable SKR of the entire network can reach $8.75$ $\rm{Gbps}$ at a uniform distance of $5$ $\rm{km}$ (asymptotic case), $0.82$ $\rm{Mbps}$ at $120$ $\rm{km}$ (finite-size effect), $89.10$ $\rm{Mbps}$ at $40$ $\rm{km}$ (compsable security), $13.66$ $\rm{Mbps}$ at $40$ $\rm{km}$ (compsable finite-size security). This achievement provides theoretical and experimental construction methods for polychromatic continuous-variable quantum communication networks, which lays the foundation for the expansion and widespread application of quantum networks.

\begin{figure*}[!ht]
\centering
\includegraphics[width=1\linewidth]{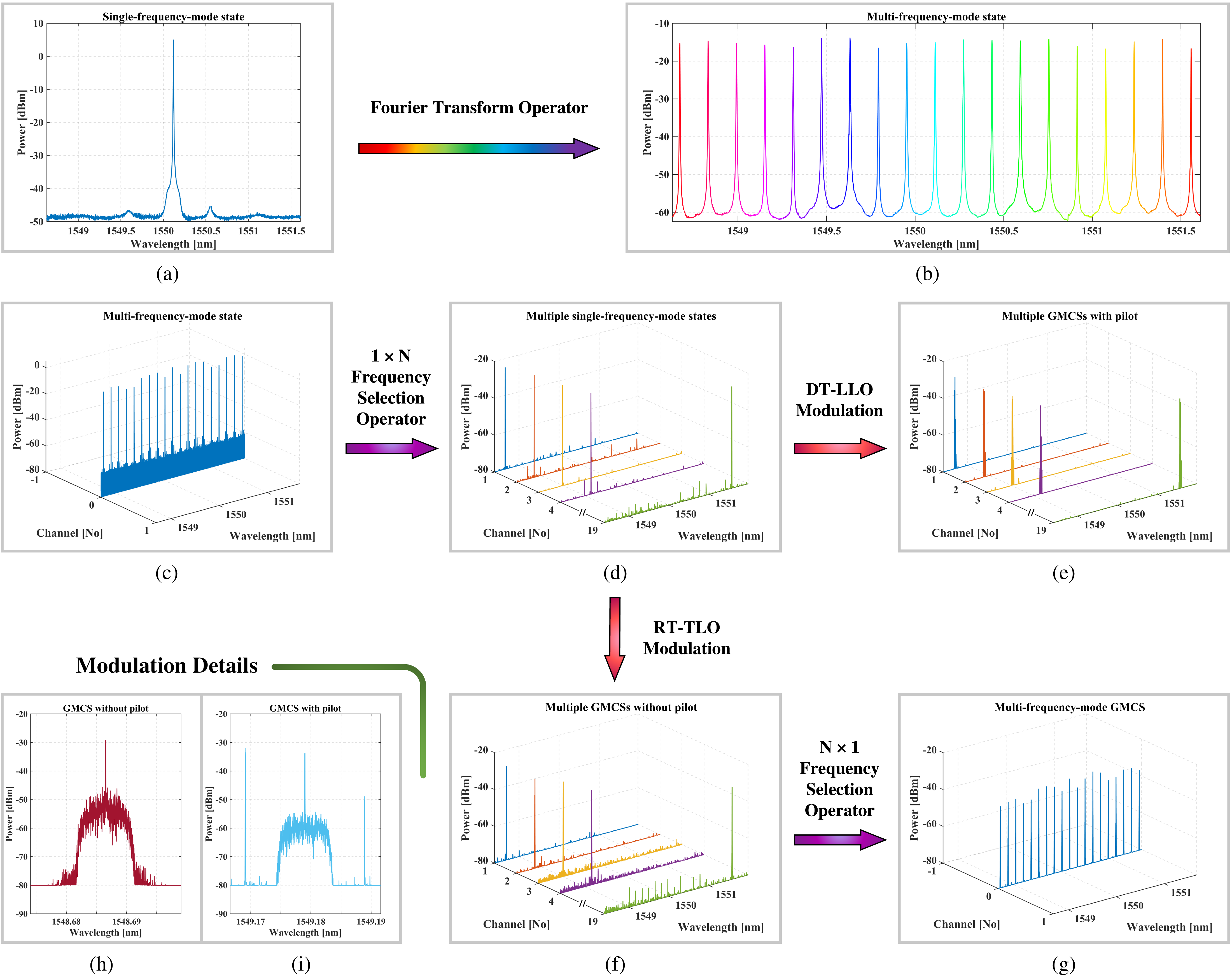}
\caption{Spectrometer measurement results in experiments. The resolution of the spectrometer used in (a) and (b) is $1.8$ $\rm{pm}$, while the others are $0.04$ $\rm{pm}$. (a) The single-mode coherent state emitted by Alice. (b) The multi-frequency-mode coherent state prepared by the generation module. (c) The multi-frequency-mode coherent state for distribution. (d) The single-frequency mode coherent states that are separated. (e) The multiple GMCSs with pilot after DT-LLO modulation. (f) The multiple GMCSs without pilot after RT-TLO modulation. (g) The multi-frequency-mode GMCS after merging. (h) The magnified GMCS without pilot of one channel. (i) The magnified GMCS with pilot of one channel.}
\label{spectrometer-result}
\end{figure*}

\section{prepare-and-measure model and entanglement-based model}

\emph{\textbf{PM model}}---The PM model describes the entire physical process of quantum states in the quantum communication network, from preparation to measurement. Physically, by utilizing polychromatic resources, this quantum network can be described as a quantum system composed of multi-frequency-mode states \cite{fabre2020modes}, and it can be used as a resource to build a star-topology network as Fig. \ref{Topology}. The maximum number of modes that can be separated from these quantum states can be defined as the network capacity, which means the maximum number of users that the entire network can accommodate (labeled as $N$ below). The PM model is described in Fig. \ref{PM}, and its specific process is as follows:

\textbf{Step 1:} Multi-frequency-mode state preparation. The central node prepares pulse signals in the time domain through a laser, forming multi-temporal-mode coherent states $\left|\left\{\gamma^\prime_0\left(t\right)\right\}\right\rangle$. Then, through the Fourier transform operator $\hat{F}_{\xi_{s,i}}$, which can be achieved through a time lens, the multi-temporal-mode coherent states can be transformed into multi-frequency-mode coherent states $\left|\left\{\gamma_m \left(\omega\right)\right\}\right\rangle$.

\textbf{Step 2:} Multi-frequency-mode state modulation. Through the frequency selection operator $\hat{M}^{\rightarrow}_{\xi_s}$, multi-frequency-mode coherent states can be separated into single-frequency-mode coherent states, and the number of separated modes is $N$. Then, one can perform Gaussian modulation on coherent states. By using the Mach-Zehnder interferometer (MZI) structure with beam splitter (BS) operator $\hat{B}_{\xi_{s,i}}$ and phase rotation operator $\hat{R}_{\xi^m_{s,i}}$, the modulation of quadrature components can be achieved, and generate coherent states $\left|x_{s,i}+jp_{s,i} \right\rangle_{\xi_{s,i}}$, with quadrature components $ x_{s,i}$, $p_{s,i}$ satisfies Gaussian distribution. Finally, the central node sends coherent states to the user nodes.

\begin{figure*}
	\centering
	\subfigure[]{\label{DT-LLO}\includegraphics[height=5.85cm]{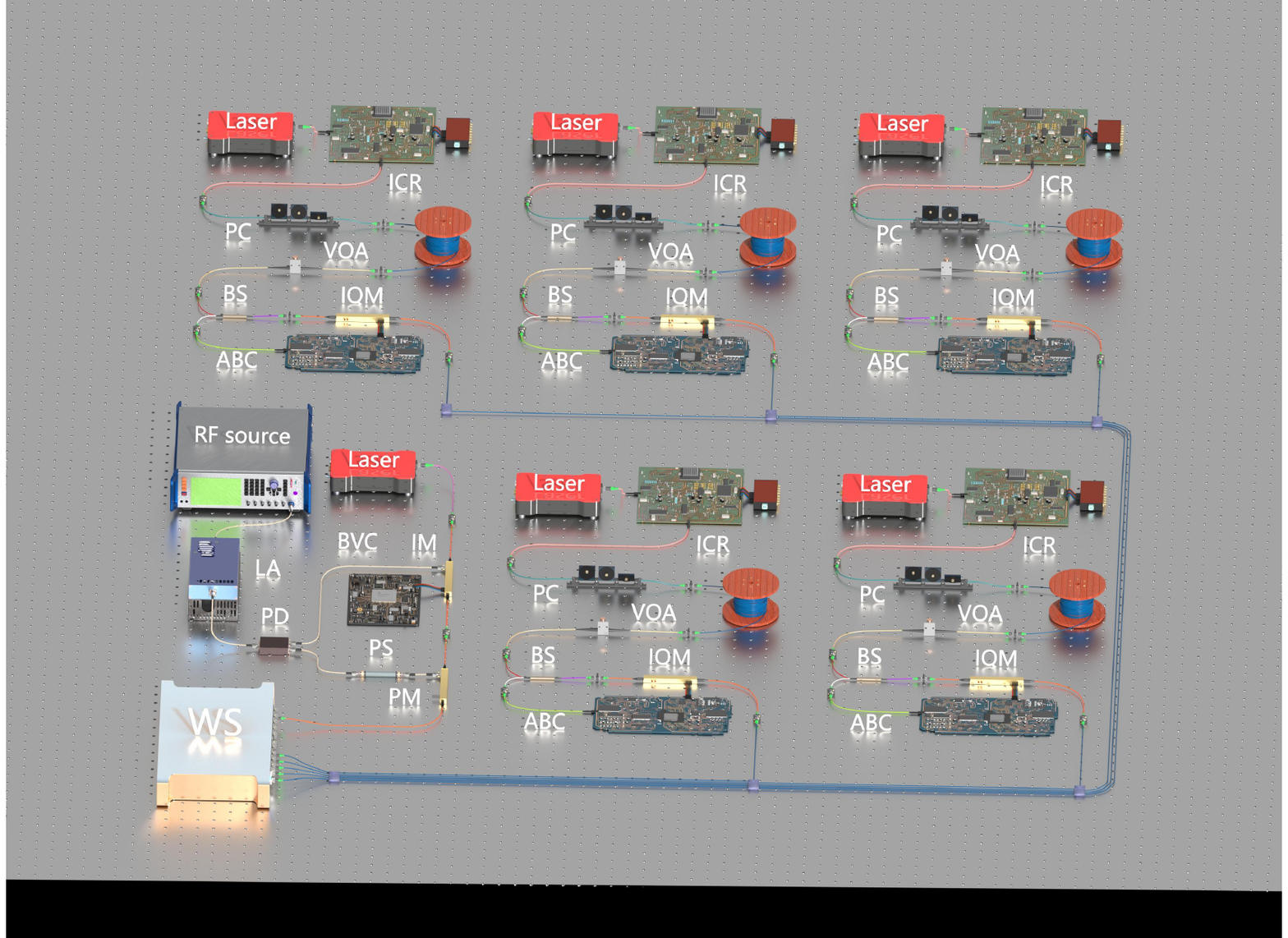}}
	\subfigure[]{\label{RT-TLO}\includegraphics[height=5.85cm]{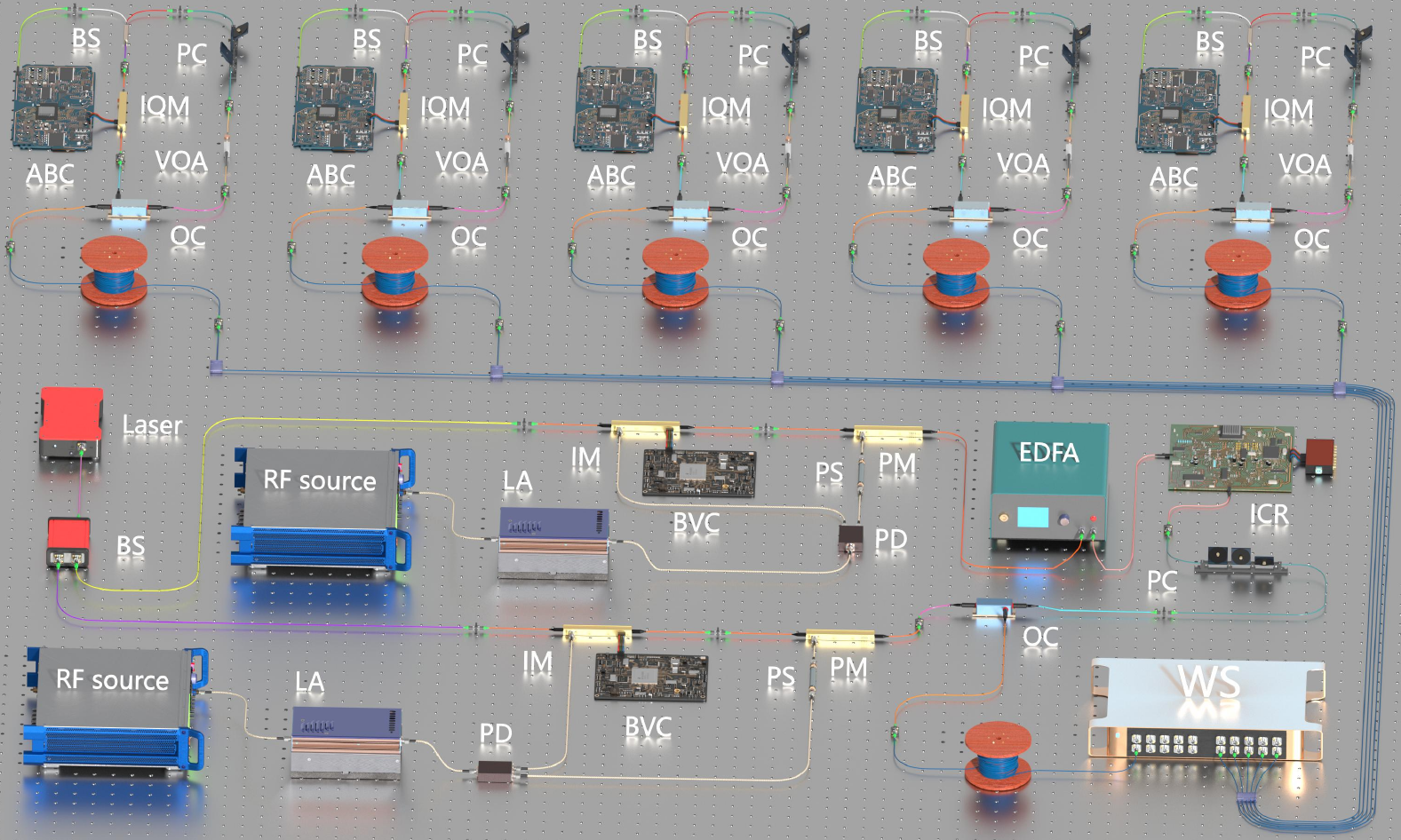}}
	\caption{The experimental set-up of DT-LLO and RT-TLO quantum network. (a) DT-LLO quantum network. The milky white braided lines are the high-frequency cables. The white fiber of $90:10$ BS corresponds to the lower-intensity output, while the red fiber corresponds to the higher-intensity output. (b) RT-TLO quantum network. The purple fiber of $99:1$ polarization-maintaining BS corresponds to the lower-intensity output, while the yellow fiber corresponds to the higher-intensity output. In OC, the pink, orange, and blue fibers are connected to ports $1$, $2$, and $3$ respectively.}
	\label{Exper-setup}
\end{figure*}

\textbf{Step 3:} Multi-frequency-mode state measurement. After channel transmission, user nodes need to detect quantum states at different frequencies. Two detection schemes are provided here. The first scheme is for each user to detect the coherent state of the single-frequency mode they receive, through coherent detection based on a single-frequency LO. The second scheme is to aggregate all modes through the inverse frequency selection operator $\hat{M}^{\leftarrow}_{\xi_s}$, and then achieve coherent detection of multi-frequency-mode states. Our subsequent experiments also demonstrated the feasibility of the two schemes separately.

\textbf{Step 4:} Post-processing. To ultimately achieve QKD, after the quantum communication process is completed, parameter estimation is required to obtain the transmittance and excess noise of each channel, to evaluate the amount of secure keys in each node. The final key is obtained through decoding and privacy amplification.

The mathematical description of the PM quantum network is presented within a time-frequency framework \cite{blow1990continuum,raymer2020temporal,fabre2020modes}. This framework not only enables the representation of quantum networks constructed using diverse multi-mode states but also streamlines the comparative analysis among different networks. For further details on this aspect, kindly refer to Notes $1-4$ provided in the Supplementary Materials.

\emph{\textbf{EB model}}---The function of the EB model is to derive the maximum amount of information that the eavesdropper can obtain. For the above PM network, entangled states of continuous variables described in Fig. \ref{EB} can be used for equivalence. Specifically, Alice prepares a multi-mode Einstein-Podolsky-Rosen (EPR) state with $N$ pairs, retaining $\rm{A}$ for detection and sending $\rm{B_0}$. Due to mutual interference between modes, $\rm{A}$ and $\rm{B_1}$ generate a fully-connected multi-mode entanglement containing $N^2$ pairs of EPR states. The entanglement degree in each pair of EPR states is represented by $\zeta^i_j$, where $i$ represents the $i$-th mode in Alice and $j$ represents the $j$-th mode in Bob. This parameter reflects the degree of crosstalk between different modes. When Bob receives $\rm{B_1}$, the multi-mode heterodyne detection introduces equivalent electronic noise for $N$ pairs of EPR states, coupled with $\rm{B_1}$ at a transmittance of the quantum detection efficiency $\eta_e$, and generated the test state $\rm{B_2}$ and the remaining states $\rm{F}$ and $\rm{G}$. It is worth noting that although all nodes are considered as Bob, they are physically separable.

Currently, there are three models for considering information capacities in CV-QKD: asymptotic case \cite{scarani2009security}, finite-size effect \cite{leverrier2010finite}, and composable security \cite{leverrier2015composable}. In addition, there is also a situation that considers both the finite-size effect and composable security simultaneously, called composable finite-size security. We extended four models of quantum networks and obtained the total SKR in the network, which can be represented as
\begin{equation}
\begin{aligned}
K= \frac{F_m k_n}{N_t} \left(\beta I_{\rm{AB}}-\chi_{\rm{BE}}-\Delta\right),
\end{aligned}
\end{equation}
where $K$ represents the total SKR, $\beta$ states the reconciliation efficiency, $I_{\rm{AB}}$ represents the Shannon mutual information in the quantum network, and $\Delta$ is related to the security of the finite-size effect and composable security. $F_m$ states the system repetition rate, $k_n$ is the coefficient of the discarded data, and $N_t$ represents the proportionality coefficient under different multiplex techniques (see Note $5$ in Supplementary Material).

The amount of information that legitimate users can obtain is determined by their respective mutual information and reconciliation efficiency, that is,
\begin{equation}
\begin{aligned}
\beta = \left[\beta_1,\beta_2,\cdots,\beta_N \right], \; I_{\rm{AB}} = \left[I^1_{\rm{AB}},I^1_{\rm{AB}},\cdots,I^N_{\rm{AB}}\right]^\intercal,
\end{aligned}
\end{equation}
in which $I^i_{\rm{AB}}$ and $\beta_i$ represents classical mutual information and the reconciliation efficiency of the $i$-th user.

\begin{figure*}[!ht]
	\centering
	\subfigure[]{\label{EN-VA}\includegraphics[width=0.32\linewidth]{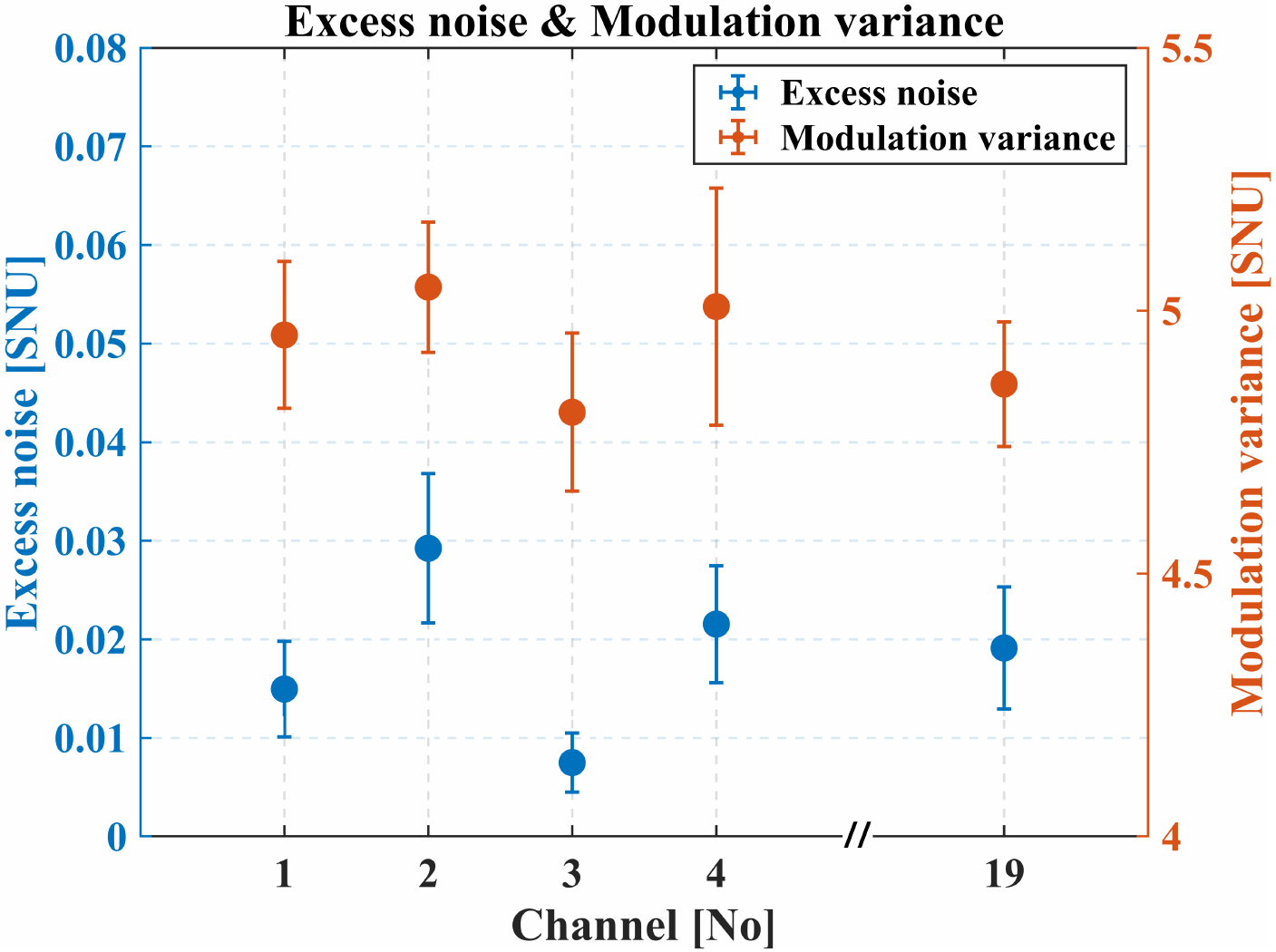}}
	\subfigure[]{\label{RE-FER}\includegraphics[width=0.32\linewidth]{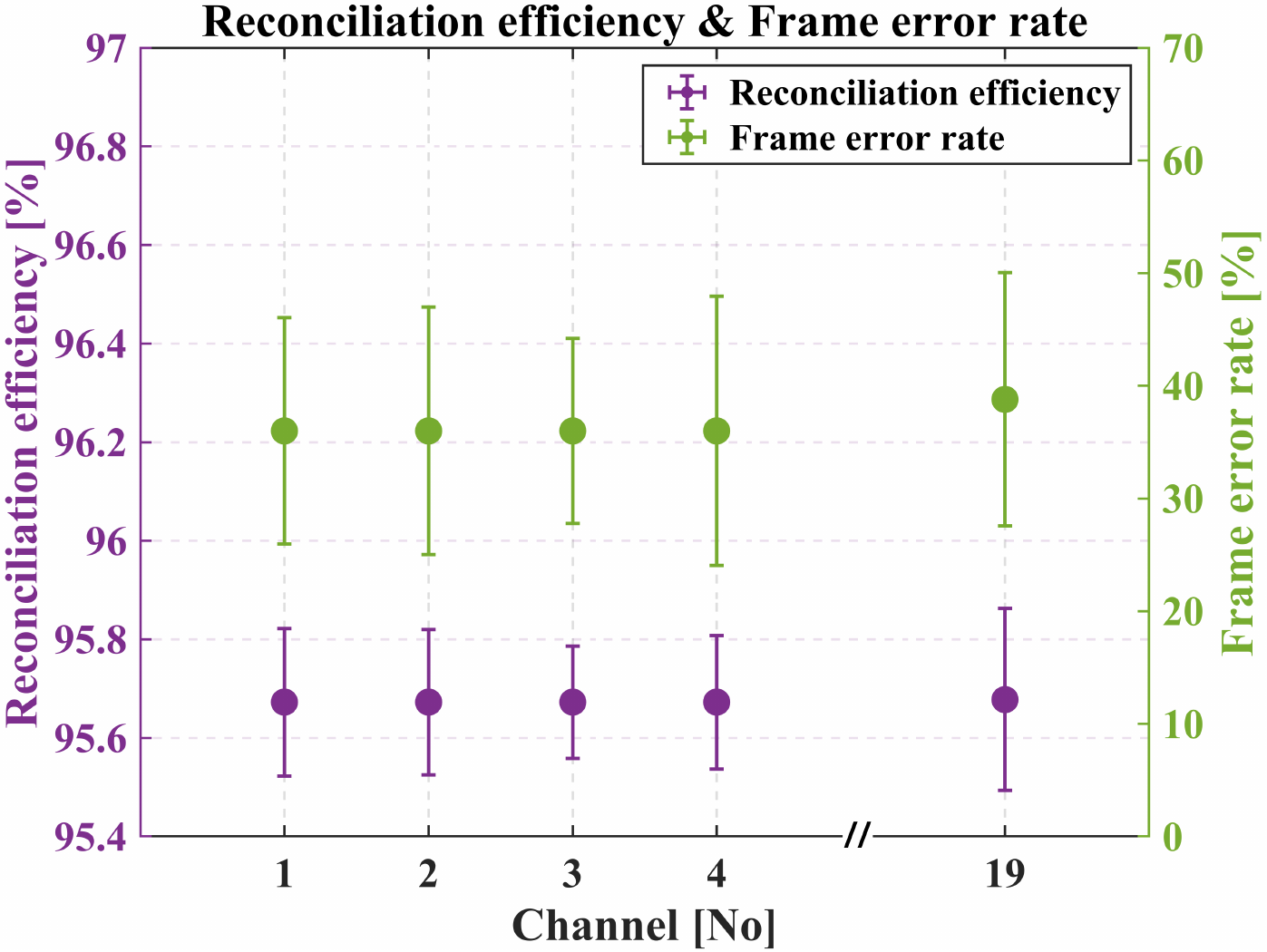}}
	\subfigure[]{\label{SKR-ex-TD}\includegraphics[width=0.32\linewidth]{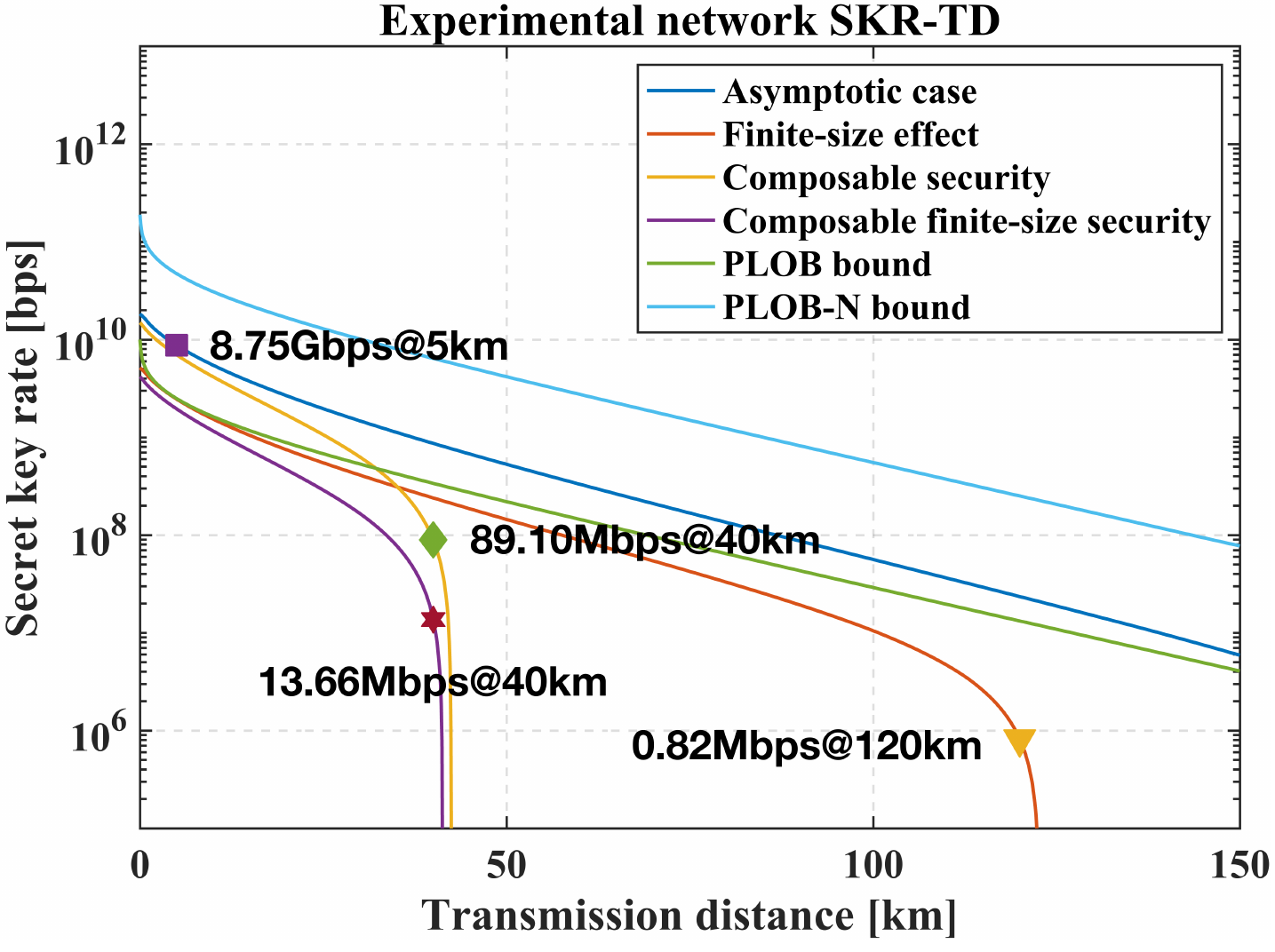}}
	\caption{Parameter estimation, decoding result, and network SKR in experiments. (a) Excess noise and modulation variance. The average of excess noise and modulation variance are $18.47$ $\rm{mSNU}$ and $4.93$ $\rm{SNU}$ respectively. (b) Reconciliation efficiency and frame error rate. The average reconciliation efficiency and frame error rates are $95.78\%$ and $43.58\%$, respectively. (c) The relationship between experimental network SKR and transmission distance. Only the asymptotic case surpassed the PLOB bound, with no experiment exceeding the PLOB-$N$ bound.}
	\label{Ex-result}
\end{figure*}

Intuitively, the total information capacity of a quantum network should be the sum of the information capacities of point-to-point links, which treats the quantum network as a linear superposition of them, However, considering the interaction between quantum states within the network, Eve in Fig. \ref{EB} can not only obtain information from its corresponding channel, but also from other channels. Therefore, we need a universal model to characterize the amount of information Eve can obtain.

To characterize $\chi_{\rm{BE}}$, we extend the Holevo bound to obtain the Holevo-$N$ bound as \cite{wolf2006extremality,garcia2006unconditional}
\begin{equation}
\label{XBE-1}
\begin{aligned}
\chi_{\rm{BE}} = S\left(\hat{\rho}_{\rm{AB_1}}\right) - S\left(\hat{\rho}^{r_{\rm{B}}}_{\rm{AFG}}\right).
\end{aligned}
\end{equation}
In a quantum network with network capacity $N$, since all modes are Gaussian states, its $N$-mode density matrix $\hat{\rho}$ can be determined by an $N$-mode covariance matrix $\Psi$ with zero mean $\hat{\rho}\left(0, \Psi \right)$ \cite{weedbrook2012gaussian}. For system $\rm{AB_1}$, the covariance matrix $\Psi_{\rm{A B_1}}$ includes $N^2$ pairs of EPR states, and describes the correlation of quadrature components between different modes. Extracting the covariance matrix between any two modes, one can obtain any pair of $N^2$ EPR states as
\begin{equation}
\label{AB1ij}
\begin{aligned}
\Psi_{\rm{A^i B^j_1}}=
\begin{bmatrix}
V_i I & \sqrt{\zeta^i_j T_d^j\left(V_i^2 -1\right)} Z \\
\sqrt{\zeta^i_j T_d^j\left(V_i^2 -1\right)} Z & T_d^j \left( \sum_{u=1}^N \zeta^u_j V_u + \chi^j_{\rm{line}} \right) I \\
\end{bmatrix},
\end{aligned}
\end{equation}
which reveals the degree of entanglement between them. For system $\rm{AFG}$ after measurement, one can also characterize it through the covariance matrix $\Psi^{r_{\rm{B}}}_{\rm{AFG}}$. By calculating the symplectic eigenvalues of the two covariance matrices above, $\left\{ \lambda_1,\lambda_2, \cdots, \lambda_{2N} \right\}$ of $\Psi_{\rm{A B_1}}$, and $\left\{ \lambda_{2N+1},\lambda_{2N+2},\cdots,\lambda_{5N} \right\}$ of $\Psi^{r_{\rm{B}}}_{\rm{A F G}}$ can be obtained, and Eq. \ref{XBE-1} can be calculated as
\begin{equation}
\label{nonlinear}
\begin{aligned}
\chi_{\rm{BE}} &= \sum_{k=1}^{2N}g\left(\lambda_k\right) - \sum_{k=2N+1}^{5N} g\left(\lambda_k\right).
\end{aligned}
\end{equation}
It is a general formula for the Holevo-$N$ bound (The detailed derivation process can refer to Note $5$ in Supplementary Materials).

\emph{\textbf{Performance}}---Based on the above information capacity calculation, we can analyze the overall performance of the quantum network. As depicted in Fig. \ref{SKR-se-NC}, \ref{SKR-se-TD}, one can see that the total SKR increases with the network capacity and decreases with distance, and is not limited by the maximum upper bound of point-to-point Pirandola-Laurenza-Ottaviani-Banchi (PLOB) bound \cite{pirandola2017fundamental,pirandola2019end,pirandola2019bounds,das2021universal}. In Fig.\ref{SKR-se-MI}, we find mode isolation greatly affects the information capacity of quantum networks, which means that better mode isolation is crucial for polychromatic quantum networks.

In addition, we also analyzed the performance of the multi-temporal-mode \cite{chen2021implementation,fan2022robust,mandil2023quantum,huang2024cost} and multi-dimension-splitter \cite{xu2023round,research.0416,qi2024experimental,pan2024high,hajomer2024continuous,li2024experimental} quantum communication networks. In Fig. \ref{SKR-ty-NC}, we find that as the number of users increases in other networks, the total SKR does not increase, but the multi-frequency-mode quantum network can continuously increase and form a pair of parallel lines with the PLOB-$N$ bound, which can represent the upper bound on the capacity of the polychromatic quantum network. The network SKR will also be higher than other networks, as shown in \ref{SKR-ty-TD}. 

High performance comes from two aspects: one is that multi-frequency modes provide sufficient bandwidth compared to multi-temporal modes, and the other is that wavelength splitting does not introduce natural channel attenuation compared with beam splitting. Since the performance of multi-frequency-mode quantum networks can satisfy $\left( K/N \right) \propto 1$, which means that under the same transmission distance, the total SKR is linear with the network capacity, we refer to this characteristic as network-capacity-independent quantum network (NCI-QN), indicating a quantum network where the single channel capacity is not related to the network capacity. This characteristic implies that one can achieve unlimited expansion of network capacity without sacrificing the key rate of a single user channel, which signifies the immense potential of polychromatic resources.

\section{Experimental set-up}

Two types of multi-frequency-mode quantum networks are set up to experimentally verify the feasibility of the NCI-QN. The first type is the DT-LLO quantum network, and the second type is the RT-TLO quantum network. The LLO and TLO schemes have been widely used in point-to-point CV-QKD. Next, we will describe these two experimental implementations separately.

\emph{\textbf{DT-LLO quantum network}}---The experimental set-up is illustrated in Fig. \ref{DT-LLO}. Initially, Alice uses an ultra-narrow-linewidth laser to emit a series of temporal-mode coherent states, which are transmitted to a generation module of multiple frequency modes realized by optical frequency combs. This module acts as a Fourier transform operator, which is composed of a radio frequency (RF) source, a linear amplifier (LA), a power divider (PD), a phase shifter (PS), a bias voltage controller (BVC), an intensity modulator (IM), and a phase modulator (PM). Specifically, the RF source generates a sine wave with $20$ $\rm{GHz}$ frequency. After the RF signal is linearly amplified, it is split into two balanced electronic signals. One part directly enters the IM, and the other part is adjusted by PS to enter the PM. After passing through the Fourier transform operator with $\ddot{\phi}_t=\sqrt{2\pi}$, the multi-temporal-mode coherent states are transformed into multi-frequency-mode coherent states with a $20$ $\rm{GHz}$ interval. By controlling BVC and PS, we can increase its mode count or make it flatter in the frequency domain, which is equivalent to changing $k$ of the Bogoliubov transformation.

Subsequently, Alice uses a $1 \times 19$ wave shaper (WS) to act as the frequency selection operator $\hat{M}^{\rightarrow}_{\xi}$ with $N=19$. By precisely adjusting the filtering parameters of the $19$ channels, it is satisfied that the selected wavepacket $\xi_{w_i}\left(\omega\right)$ correspond to each wavepacket $\xi_{s,i}\left(\omega\right)$ of multi-frequency-mode coherent states in $\hat{S}^1_{\xi}$ and $\hat{S}^2_{\xi}$, separating multiple single-frequency-mode coherent states. This process also affects the mode isolation. At this point, the highest optical power difference between the adjacent modes and the main mode is defined as mode isolation. Eventually, this frequency selection operator can achieve a mode isolation of less than $-50$ $\rm{dB}$. In this case, the influence of adjacent frequency modes on the quantum network can be virtually negligible.

The individual single-frequency-mode coherent states need to be modulated separately. Each modulation module consists of an in-phase and quadrature modulator (IQM), a BS with a splitting ratio of $90:10$, an automatic bias controller (ABC), and a variable optical attenuator (VOA). Every coherent state enters the IQM for independent baseband Gaussian modulation. To achieve a nearly perfect evolution, the modulated coherent state passes through the $90:10$ BS to split off some power into ABC. It adjusts the bias of the IQM in real time to minimize the impact of imbalance. Finally, the coherent states are attenuated by a VOA, resulting in Gaussian-modulated coherent states (GMCS). Alice has already prepared $19$ sets of GMCS.

Concerning the channel, multiple independent GMCSs are transmitted through multiple physical channels. For multiple sets of experiments, we used optical fiber cables of various lengths. The $i$-th state output by the channel is $\left| \gamma_{1,i} \right\rangle_{\xi_{s,i}}$.

Bob uses multiple single-mode receivers to receive these quantum states. Each single-mode receiver consists of a polarization controller (PC), an ultra-narrow-linewidth laser, and an integrated coherent detector (ICR). The manual PC is used to adjust the polarization of the received GMCS to align with the polarization of the LO. The GMCS and LO are input into the ICR for heterodyne detection. After coherent monochromatic detection, Bob needs to perform DSP on the data to achieve optimal reception \cite{chen2023continuous,xu2024robust}.

\emph{\textbf{RT-TLO quantum network}}---The experimental set-up is illustrated in Fig. \ref{RT-TLO}. In the round-trip structure, Alice is responsible for generating and receiving coherent states, while Bob only needs to modulate the coherent states. Firstly, Alice's ultra-narrow-linewidth laser is split into two beams by a $99:1$ polarization-maintaining BS. The higher-intensity part passes through LO's multi-frequency-mode generation module and then enters an erbium-doped fiber amplifier (EDFA), while the lower-intensity part goes into the multi-frequency-mode generation module of a coherent state. The former one operates at $19$ $\rm{GHz}$, while the latter one operates at $20$ $\rm{GHz}$. Subsequently, it enters from port $1$ of the optical circulator (OC), and exits from port $2$ to the channel.

The round-trip network transmits a multi-frequency-mode coherent state in the channel. Specifically, the quantum channel consists of a shared optical fiber cable, a WS, and several optional optical fiber cables. The multi-frequency-mode coherent state first enters the shared optical fiber cable for transmission. In the direction from Alice to Bob, WS functions to separate the multi-mode coherent state as the frequency selection operator $\hat{M}^{\rightarrow}_{\xi}$. The single-mode coherent states then pass through their respective optional optical fiber cables to reach Bob. Bob receives the coherent state at port $2$ of the OC, and then outputs it from port $3$ to the modulation module. The modulation process of IQM is similar to that of the DT-LLO quantum network, except for one more PC. After modulation, the GMCS enters port $1$ of the OC and then exits from port $2$ into the channel.

In the direction from Bob to Alice, a multi-frequency-mode GMCS is transmitted in the channel. Multiple independent GMCSs first enter their respective optional optical fiber cables for transmission. Then, they enter the WS where they converge into a multi-frequency-mode GMCS. Here, the role of the WS is to serve as the frequency selection operator $\hat{M}^{\leftarrow}_{\xi}$. The expression of the multi-mode GMCS output from a shared optical fiber cable can be indicated by $\left| \left\{ \gamma_1 (\omega) \right\} \right\rangle$. Ultimately, it returns to port $2$ of Alice's OC and is transmitted from port $3$ to the PC. After the polarization is adjusted once again through the PC, multi-mode GMCS enters the ICR for heterodyne detection. Alice constructs a multi-mode receiver with a multi-mode LO to detect the multi-mode quantum state. After coherent polychromic detection, Alice needs to perform $N$ DSPs on the data to achieve optimal reception. The experimental device parameters and detailed processes are shown in Note $6$ of Supplementary Materials.

\section{Experimental results}

The spectrometer measurement results are depicted in Fig. \ref{spectrometer-result}. The parameter estimation, decoding result, and network SKR are shown in Fig. \ref{Ex-result}. When evaluating signal noise, shot noise, and electronic noise, the same DSP methods are applied to the same channel to avoid security issues. The decoding method used in the experiments is multi-edge-type low-density parity-check (MET-LDPC) codes. Four types of experiments are conducted, including asymptotic case, finite-size effect, composable security, and composable finite-size security. In the asymptotic case, the network SKR is $8.75$ $\rm{Gbps}$ at $5$ $\rm{km}$. Under finite-size effect, the network SKR is $0.82$ $\rm{Mbps}$ at $120$ $\rm{km}$. With composable security, the network SKR is $89.10$ $\rm{Mbps}$ at $40$ $\rm{km}$. For the composable finite-size security, the network SKR is $13.66$ $\rm{Mbps}$ at $40$ $\rm{km}$. Specific experimental data can be found in Note $7$ of the Supplementary Materials. In the above experiment, one can see that the SKR for each user is independent of network capacity, which only depends on the number of modes at the beginning of preparing the quantum state. Therefore, the information capacity per user-channel is independent of the network capacity $\left( K/N \right) \propto 1$ has been proved experimentally.

\section{Conclusion}

In this letter, we have presented a scheme for the implementation of a polychromatic continuous-variable quantum network, which has the characteristic of NCI-QN, and is verified through both theoretical and experimental aspects. We formulate the total SKR in our network scheme according to the EB model. Besides, the reachable SKR is experimentally obtained under conditions of the asymptotic case, finite-size effect, composable security, and composable finite-size security, and the total SKR in the network can reach up to $8.75$ $\rm{Gbps}$. All elements of this scheme fall within the scope of current experimental technologies and exhibit excellent scalability. It allows for the construction of a quantum network wherein the information capacity per user channel remains invariant to the increase in the users. This high-performance quantum network, combined with the experimental setup compatible with existing optical networks, offers strong support for the future development of the quantum internet.

\section*{Acknowledgments}

G. Z. conceived the research. Y. X., Q. Z., J. Z. and X. L. proposed the physical models and completed mathematical proofs. Y. X., X. L., Z. S., X. L., B. Z., Z. T., Z. Z. and T. W. constructed the network platform and conducted experimental verification. J. D. and X. J. provided the post-processing decoding program. G. Z., T. W. and P. H. guided the calculation of information capacity. All authors together wrote and reviewed this manuscript.

This work is supported by Innovation Program for Quantum Science and Technology (Grant No. 2021ZD0300703), National Natural Science Foundation of China (Grant No. 62101320), Shanghai Municipal Science and Technology Major Project (Grant No. 2019SHZDZX01).

\bibliography{NCI-QN}

\begin{thebibliography}{73}%
\makeatletter
\providecommand \@ifxundefined [1]{%
 \@ifx{#1\undefined}
}%
\providecommand \@ifnum [1]{%
 \ifnum #1\expandafter \@firstoftwo
 \else \expandafter \@secondoftwo
 \fi
}%
\providecommand \@ifx [1]{%
 \ifx #1\expandafter \@firstoftwo
 \else \expandafter \@secondoftwo
 \fi
}%
\providecommand \natexlab [1]{#1}%
\providecommand \enquote  [1]{``#1''}%
\providecommand \bibnamefont  [1]{#1}%
\providecommand \bibfnamefont [1]{#1}%
\providecommand \citenamefont [1]{#1}%
\providecommand \href@noop [0]{\@secondoftwo}%
\providecommand \href [0]{\begingroup \@sanitize@url \@href}%
\providecommand \@href[1]{\@@startlink{#1}\@@href}%
\providecommand \@@href[1]{\endgroup#1\@@endlink}%
\providecommand \@sanitize@url [0]{\catcode `\\12\catcode `\$12\catcode
  `\&12\catcode `\#12\catcode `\^12\catcode `\_12\catcode `\%12\relax}%
\providecommand \@@startlink[1]{}%
\providecommand \@@endlink[0]{}%
\providecommand \url  [0]{\begingroup\@sanitize@url \@url }%
\providecommand \@url [1]{\endgroup\@href {#1}{\urlprefix }}%
\providecommand \urlprefix  [0]{URL }%
\providecommand \Eprint [0]{\href }%
\providecommand \doibase [0]{https://doi.org/}%
\providecommand \selectlanguage [0]{\@gobble}%
\providecommand \bibinfo  [0]{\@secondoftwo}%
\providecommand \bibfield  [0]{\@secondoftwo}%
\providecommand \translation [1]{[#1]}%
\providecommand \BibitemOpen [0]{}%
\providecommand \bibitemStop [0]{}%
\providecommand \bibitemNoStop [0]{.\EOS\space}%
\providecommand \EOS [0]{\spacefactor3000\relax}%
\providecommand \BibitemShut  [1]{\csname bibitem#1\endcsname}%
\let\auto@bib@innerbib\@empty
\bibitem [{\citenamefont {Townsend}(1997)}]{townsend1997quantum}%
  \BibitemOpen
  \bibfield  {author} {\bibinfo {author} {\bibfnamefont {P.~D.}\ \bibnamefont
  {Townsend}},\ }\bibfield  {title} {\bibinfo {title} {Quantum cryptography on
  multiuser optical fibre networks},\ }\href@noop {} {\bibfield  {journal}
  {\bibinfo  {journal} {Nature}\ }\textbf {\bibinfo {volume} {385}},\ \bibinfo
  {pages} {47} (\bibinfo {year} {1997})}\BibitemShut {NoStop}%
\bibitem [{\citenamefont {Duan}\ \emph {et~al.}(2001)\citenamefont {Duan},
  \citenamefont {Lukin}, \citenamefont {Cirac},\ and\ \citenamefont
  {Zoller}}]{duan2001long}%
  \BibitemOpen
  \bibfield  {author} {\bibinfo {author} {\bibfnamefont {L.-M.}\ \bibnamefont
  {Duan}}, \bibinfo {author} {\bibfnamefont {M.~D.}\ \bibnamefont {Lukin}},
  \bibinfo {author} {\bibfnamefont {J.~I.}\ \bibnamefont {Cirac}},\ and\
  \bibinfo {author} {\bibfnamefont {P.}~\bibnamefont {Zoller}},\ }\bibfield
  {title} {\bibinfo {title} {Long-distance quantum communication with atomic
  ensembles and linear optics},\ }\href@noop {} {\bibfield  {journal} {\bibinfo
   {journal} {Nature}\ }\textbf {\bibinfo {volume} {414}},\ \bibinfo {pages}
  {413} (\bibinfo {year} {2001})}\BibitemShut {NoStop}%
\bibitem [{\citenamefont {Choi}\ \emph {et~al.}(2011)\citenamefont {Choi},
  \citenamefont {Young},\ and\ \citenamefont {Townsend}}]{choi2011quantum}%
  \BibitemOpen
  \bibfield  {author} {\bibinfo {author} {\bibfnamefont {I.}~\bibnamefont
  {Choi}}, \bibinfo {author} {\bibfnamefont {R.~J.}\ \bibnamefont {Young}},\
  and\ \bibinfo {author} {\bibfnamefont {P.~D.}\ \bibnamefont {Townsend}},\
  }\bibfield  {title} {\bibinfo {title} {Quantum information to the home},\
  }\href@noop {} {\bibfield  {journal} {\bibinfo  {journal} {New Journal of
  Physics}\ }\textbf {\bibinfo {volume} {13}},\ \bibinfo {pages} {063039}
  (\bibinfo {year} {2011})}\BibitemShut {NoStop}%
\bibitem [{\citenamefont {Fr{\"o}hlich}\ \emph {et~al.}(2013)\citenamefont
  {Fr{\"o}hlich}, \citenamefont {Dynes}, \citenamefont {Lucamarini},
  \citenamefont {Sharpe}, \citenamefont {Yuan},\ and\ \citenamefont
  {Shields}}]{frohlich2013quantum}%
  \BibitemOpen
  \bibfield  {author} {\bibinfo {author} {\bibfnamefont {B.}~\bibnamefont
  {Fr{\"o}hlich}}, \bibinfo {author} {\bibfnamefont {J.~F.}\ \bibnamefont
  {Dynes}}, \bibinfo {author} {\bibfnamefont {M.}~\bibnamefont {Lucamarini}},
  \bibinfo {author} {\bibfnamefont {A.~W.}\ \bibnamefont {Sharpe}}, \bibinfo
  {author} {\bibfnamefont {Z.}~\bibnamefont {Yuan}},\ and\ \bibinfo {author}
  {\bibfnamefont {A.~J.}\ \bibnamefont {Shields}},\ }\bibfield  {title}
  {\bibinfo {title} {A quantum access network},\ }\href@noop {} {\bibfield
  {journal} {\bibinfo  {journal} {Nature}\ }\textbf {\bibinfo {volume} {501}},\
  \bibinfo {pages} {69} (\bibinfo {year} {2013})}\BibitemShut {NoStop}%
\bibitem [{\citenamefont {Fr{\"o}hlich}\ \emph {et~al.}(2015)\citenamefont
  {Fr{\"o}hlich}, \citenamefont {Dynes}, \citenamefont {Lucamarini},
  \citenamefont {Sharpe}, \citenamefont {Tam}, \citenamefont {Yuan},\ and\
  \citenamefont {Shields}}]{frohlich2015quantum}%
  \BibitemOpen
  \bibfield  {author} {\bibinfo {author} {\bibfnamefont {B.}~\bibnamefont
  {Fr{\"o}hlich}}, \bibinfo {author} {\bibfnamefont {J.~F.}\ \bibnamefont
  {Dynes}}, \bibinfo {author} {\bibfnamefont {M.}~\bibnamefont {Lucamarini}},
  \bibinfo {author} {\bibfnamefont {A.~W.}\ \bibnamefont {Sharpe}}, \bibinfo
  {author} {\bibfnamefont {S.~W.-B.}\ \bibnamefont {Tam}}, \bibinfo {author}
  {\bibfnamefont {Z.}~\bibnamefont {Yuan}},\ and\ \bibinfo {author}
  {\bibfnamefont {A.~J.}\ \bibnamefont {Shields}},\ }\bibfield  {title}
  {\bibinfo {title} {Quantum secured gigabit optical access networks},\
  }\href@noop {} {\bibfield  {journal} {\bibinfo  {journal} {Scientific
  reports}\ }\textbf {\bibinfo {volume} {5}},\ \bibinfo {pages} {18121}
  (\bibinfo {year} {2015})}\BibitemShut {NoStop}%
\bibitem [{\citenamefont {Liao}\ \emph {et~al.}(2017)\citenamefont {Liao},
  \citenamefont {Cai}, \citenamefont {Liu}, \citenamefont {Zhang},
  \citenamefont {Li}, \citenamefont {Ren}, \citenamefont {Yin}, \citenamefont
  {Shen}, \citenamefont {Cao}, \citenamefont {Li} \emph
  {et~al.}}]{liao2017satellite}%
  \BibitemOpen
  \bibfield  {author} {\bibinfo {author} {\bibfnamefont {S.-K.}\ \bibnamefont
  {Liao}}, \bibinfo {author} {\bibfnamefont {W.-Q.}\ \bibnamefont {Cai}},
  \bibinfo {author} {\bibfnamefont {W.-Y.}\ \bibnamefont {Liu}}, \bibinfo
  {author} {\bibfnamefont {L.}~\bibnamefont {Zhang}}, \bibinfo {author}
  {\bibfnamefont {Y.}~\bibnamefont {Li}}, \bibinfo {author} {\bibfnamefont
  {J.-G.}\ \bibnamefont {Ren}}, \bibinfo {author} {\bibfnamefont
  {J.}~\bibnamefont {Yin}}, \bibinfo {author} {\bibfnamefont {Q.}~\bibnamefont
  {Shen}}, \bibinfo {author} {\bibfnamefont {Y.}~\bibnamefont {Cao}}, \bibinfo
  {author} {\bibfnamefont {Z.-P.}\ \bibnamefont {Li}}, \emph {et~al.},\
  }\bibfield  {title} {\bibinfo {title} {Satellite-to-ground quantum key
  distribution},\ }\href@noop {} {\bibfield  {journal} {\bibinfo  {journal}
  {Nature}\ }\textbf {\bibinfo {volume} {549}},\ \bibinfo {pages} {43}
  (\bibinfo {year} {2017})}\BibitemShut {NoStop}%
\bibitem [{\citenamefont {Wengerowsky}\ \emph {et~al.}(2018)\citenamefont
  {Wengerowsky}, \citenamefont {Joshi}, \citenamefont {Steinlechner},
  \citenamefont {H{\"u}bel},\ and\ \citenamefont
  {Ursin}}]{wengerowsky2018entanglement}%
  \BibitemOpen
  \bibfield  {author} {\bibinfo {author} {\bibfnamefont {S.}~\bibnamefont
  {Wengerowsky}}, \bibinfo {author} {\bibfnamefont {S.~K.}\ \bibnamefont
  {Joshi}}, \bibinfo {author} {\bibfnamefont {F.}~\bibnamefont {Steinlechner}},
  \bibinfo {author} {\bibfnamefont {H.}~\bibnamefont {H{\"u}bel}},\ and\
  \bibinfo {author} {\bibfnamefont {R.}~\bibnamefont {Ursin}},\ }\bibfield
  {title} {\bibinfo {title} {An entanglement-based wavelength-multiplexed
  quantum communication network},\ }\href@noop {} {\bibfield  {journal}
  {\bibinfo  {journal} {Nature}\ }\textbf {\bibinfo {volume} {564}},\ \bibinfo
  {pages} {225} (\bibinfo {year} {2018})}\BibitemShut {NoStop}%
\bibitem [{\citenamefont {Dynes}\ \emph {et~al.}(2019)\citenamefont {Dynes},
  \citenamefont {Wonfor}, \citenamefont {Tam}, \citenamefont {Sharpe},
  \citenamefont {Takahashi}, \citenamefont {Lucamarini}, \citenamefont {Plews},
  \citenamefont {Yuan}, \citenamefont {Dixon}, \citenamefont {Cho} \emph
  {et~al.}}]{dynes2019cambridge}%
  \BibitemOpen
  \bibfield  {author} {\bibinfo {author} {\bibfnamefont {J.}~\bibnamefont
  {Dynes}}, \bibinfo {author} {\bibfnamefont {A.}~\bibnamefont {Wonfor}},
  \bibinfo {author} {\bibfnamefont {W.-S.}\ \bibnamefont {Tam}}, \bibinfo
  {author} {\bibfnamefont {A.}~\bibnamefont {Sharpe}}, \bibinfo {author}
  {\bibfnamefont {R.}~\bibnamefont {Takahashi}}, \bibinfo {author}
  {\bibfnamefont {M.}~\bibnamefont {Lucamarini}}, \bibinfo {author}
  {\bibfnamefont {A.}~\bibnamefont {Plews}}, \bibinfo {author} {\bibfnamefont
  {Z.}~\bibnamefont {Yuan}}, \bibinfo {author} {\bibfnamefont {A.}~\bibnamefont
  {Dixon}}, \bibinfo {author} {\bibfnamefont {J.}~\bibnamefont {Cho}}, \emph
  {et~al.},\ }\bibfield  {title} {\bibinfo {title} {Cambridge quantum
  network},\ }\href@noop {} {\bibfield  {journal} {\bibinfo  {journal} {npj
  Quantum Information}\ }\textbf {\bibinfo {volume} {5}},\ \bibinfo {pages}
  {101} (\bibinfo {year} {2019})}\BibitemShut {NoStop}%
\bibitem [{\citenamefont {Joshi}\ \emph {et~al.}(2020)\citenamefont {Joshi},
  \citenamefont {Aktas}, \citenamefont {Wengerowsky}, \citenamefont
  {Lon{\v{c}}ari{\'c}}, \citenamefont {Neumann}, \citenamefont {Liu},
  \citenamefont {Scheidl}, \citenamefont {Lorenzo}, \citenamefont {Samec},
  \citenamefont {Kling} \emph {et~al.}}]{joshi2020trusted}%
  \BibitemOpen
  \bibfield  {author} {\bibinfo {author} {\bibfnamefont {S.~K.}\ \bibnamefont
  {Joshi}}, \bibinfo {author} {\bibfnamefont {D.}~\bibnamefont {Aktas}},
  \bibinfo {author} {\bibfnamefont {S.}~\bibnamefont {Wengerowsky}}, \bibinfo
  {author} {\bibfnamefont {M.}~\bibnamefont {Lon{\v{c}}ari{\'c}}}, \bibinfo
  {author} {\bibfnamefont {S.~P.}\ \bibnamefont {Neumann}}, \bibinfo {author}
  {\bibfnamefont {B.}~\bibnamefont {Liu}}, \bibinfo {author} {\bibfnamefont
  {T.}~\bibnamefont {Scheidl}}, \bibinfo {author} {\bibfnamefont {G.~C.}\
  \bibnamefont {Lorenzo}}, \bibinfo {author} {\bibfnamefont
  {{\v{Z}}.}~\bibnamefont {Samec}}, \bibinfo {author} {\bibfnamefont
  {L.}~\bibnamefont {Kling}}, \emph {et~al.},\ }\bibfield  {title} {\bibinfo
  {title} {A trusted node--free eight-user metropolitan quantum communication
  network},\ }\href@noop {} {\bibfield  {journal} {\bibinfo  {journal} {Science
  advances}\ }\textbf {\bibinfo {volume} {6}},\ \bibinfo {pages} {eaba0959}
  (\bibinfo {year} {2020})}\BibitemShut {NoStop}%
\bibitem [{\citenamefont {Chen}\ \emph
  {et~al.}(2021{\natexlab{a}})\citenamefont {Chen}, \citenamefont {Zhang},
  \citenamefont {Chen}, \citenamefont {Cai}, \citenamefont {Liao},
  \citenamefont {Zhang}, \citenamefont {Chen}, \citenamefont {Yin},
  \citenamefont {Ren}, \citenamefont {Chen} \emph
  {et~al.}}]{chen2021integrated}%
  \BibitemOpen
  \bibfield  {author} {\bibinfo {author} {\bibfnamefont {Y.-A.}\ \bibnamefont
  {Chen}}, \bibinfo {author} {\bibfnamefont {Q.}~\bibnamefont {Zhang}},
  \bibinfo {author} {\bibfnamefont {T.-Y.}\ \bibnamefont {Chen}}, \bibinfo
  {author} {\bibfnamefont {W.-Q.}\ \bibnamefont {Cai}}, \bibinfo {author}
  {\bibfnamefont {S.-K.}\ \bibnamefont {Liao}}, \bibinfo {author}
  {\bibfnamefont {J.}~\bibnamefont {Zhang}}, \bibinfo {author} {\bibfnamefont
  {K.}~\bibnamefont {Chen}}, \bibinfo {author} {\bibfnamefont {J.}~\bibnamefont
  {Yin}}, \bibinfo {author} {\bibfnamefont {J.-G.}\ \bibnamefont {Ren}},
  \bibinfo {author} {\bibfnamefont {Z.}~\bibnamefont {Chen}}, \emph {et~al.},\
  }\bibfield  {title} {\bibinfo {title} {An integrated space-to-ground quantum
  communication network over 4,600 kilometres},\ }\href@noop {} {\bibfield
  {journal} {\bibinfo  {journal} {Nature}\ }\textbf {\bibinfo {volume} {589}},\
  \bibinfo {pages} {214} (\bibinfo {year} {2021}{\natexlab{a}})}\BibitemShut
  {NoStop}%
\bibitem [{\citenamefont {Qi}\ \emph {et~al.}(2021)\citenamefont {Qi},
  \citenamefont {Li}, \citenamefont {Huang}, \citenamefont {Feng},
  \citenamefont {Zheng},\ and\ \citenamefont {Chen}}]{qi202115}%
  \BibitemOpen
  \bibfield  {author} {\bibinfo {author} {\bibfnamefont {Z.}~\bibnamefont
  {Qi}}, \bibinfo {author} {\bibfnamefont {Y.}~\bibnamefont {Li}}, \bibinfo
  {author} {\bibfnamefont {Y.}~\bibnamefont {Huang}}, \bibinfo {author}
  {\bibfnamefont {J.}~\bibnamefont {Feng}}, \bibinfo {author} {\bibfnamefont
  {Y.}~\bibnamefont {Zheng}},\ and\ \bibinfo {author} {\bibfnamefont
  {X.}~\bibnamefont {Chen}},\ }\bibfield  {title} {\bibinfo {title} {A 15-user
  quantum secure direct communication network},\ }\href@noop {} {\bibfield
  {journal} {\bibinfo  {journal} {Light: Science \& Applications}\ }\textbf
  {\bibinfo {volume} {10}},\ \bibinfo {pages} {183} (\bibinfo {year}
  {2021})}\BibitemShut {NoStop}%
\bibitem [{\citenamefont {Chen}\ \emph
  {et~al.}(2021{\natexlab{b}})\citenamefont {Chen}, \citenamefont {Jiang},
  \citenamefont {Tang}, \citenamefont {Zhou}, \citenamefont {Yuan},
  \citenamefont {Zhou}, \citenamefont {Wang}, \citenamefont {Liu},
  \citenamefont {Chen}, \citenamefont {Liu} \emph
  {et~al.}}]{chen2021implementation}%
  \BibitemOpen
  \bibfield  {author} {\bibinfo {author} {\bibfnamefont {T.-Y.}\ \bibnamefont
  {Chen}}, \bibinfo {author} {\bibfnamefont {X.}~\bibnamefont {Jiang}},
  \bibinfo {author} {\bibfnamefont {S.-B.}\ \bibnamefont {Tang}}, \bibinfo
  {author} {\bibfnamefont {L.}~\bibnamefont {Zhou}}, \bibinfo {author}
  {\bibfnamefont {X.}~\bibnamefont {Yuan}}, \bibinfo {author} {\bibfnamefont
  {H.}~\bibnamefont {Zhou}}, \bibinfo {author} {\bibfnamefont {J.}~\bibnamefont
  {Wang}}, \bibinfo {author} {\bibfnamefont {Y.}~\bibnamefont {Liu}}, \bibinfo
  {author} {\bibfnamefont {L.-K.}\ \bibnamefont {Chen}}, \bibinfo {author}
  {\bibfnamefont {W.-Y.}\ \bibnamefont {Liu}}, \emph {et~al.},\ }\bibfield
  {title} {\bibinfo {title} {Implementation of a 46-node quantum metropolitan
  area network},\ }\href@noop {} {\bibfield  {journal} {\bibinfo  {journal}
  {npj Quantum Information}\ }\textbf {\bibinfo {volume} {7}},\ \bibinfo
  {pages} {134} (\bibinfo {year} {2021}{\natexlab{b}})}\BibitemShut {NoStop}%
\bibitem [{\citenamefont {Huang}\ \emph {et~al.}(2021)\citenamefont {Huang},
  \citenamefont {Shen}, \citenamefont {Wang}, \citenamefont {Chen},
  \citenamefont {Xu}, \citenamefont {Yu},\ and\ \citenamefont
  {Guo}}]{huang2021realizing}%
  \BibitemOpen
  \bibfield  {author} {\bibinfo {author} {\bibfnamefont {Y.}~\bibnamefont
  {Huang}}, \bibinfo {author} {\bibfnamefont {T.}~\bibnamefont {Shen}},
  \bibinfo {author} {\bibfnamefont {X.}~\bibnamefont {Wang}}, \bibinfo {author}
  {\bibfnamefont {Z.}~\bibnamefont {Chen}}, \bibinfo {author} {\bibfnamefont
  {B.}~\bibnamefont {Xu}}, \bibinfo {author} {\bibfnamefont {S.}~\bibnamefont
  {Yu}},\ and\ \bibinfo {author} {\bibfnamefont {H.}~\bibnamefont {Guo}},\
  }\bibfield  {title} {\bibinfo {title} {Realizing a downstream-access network
  using continuous-variable quantum key distribution},\ }\href@noop {}
  {\bibfield  {journal} {\bibinfo  {journal} {Physical Review Applied}\
  }\textbf {\bibinfo {volume} {16}},\ \bibinfo {pages} {064051} (\bibinfo
  {year} {2021})}\BibitemShut {NoStop}%
\bibitem [{\citenamefont {Wang}\ \emph {et~al.}(2021)\citenamefont {Wang},
  \citenamefont {Tang}, \citenamefont {Mao}, \citenamefont {Xu}, \citenamefont
  {Cheng}, \citenamefont {Zhang}, \citenamefont {Chen},\ and\ \citenamefont
  {Pan}}]{wang2021practical}%
  \BibitemOpen
  \bibfield  {author} {\bibinfo {author} {\bibfnamefont {B.-X.}\ \bibnamefont
  {Wang}}, \bibinfo {author} {\bibfnamefont {S.-B.}\ \bibnamefont {Tang}},
  \bibinfo {author} {\bibfnamefont {Y.}~\bibnamefont {Mao}}, \bibinfo {author}
  {\bibfnamefont {W.}~\bibnamefont {Xu}}, \bibinfo {author} {\bibfnamefont
  {M.}~\bibnamefont {Cheng}}, \bibinfo {author} {\bibfnamefont
  {J.}~\bibnamefont {Zhang}}, \bibinfo {author} {\bibfnamefont {T.-Y.}\
  \bibnamefont {Chen}},\ and\ \bibinfo {author} {\bibfnamefont {J.-W.}\
  \bibnamefont {Pan}},\ }\bibfield  {title} {\bibinfo {title} {Practical
  quantum access network over a 10 gbit/s ethernet passive optical network},\
  }\href@noop {} {\bibfield  {journal} {\bibinfo  {journal} {Optics Express}\
  }\textbf {\bibinfo {volume} {29}},\ \bibinfo {pages} {38582} (\bibinfo {year}
  {2021})}\BibitemShut {NoStop}%
\bibitem [{\citenamefont {Wang}\ \emph {et~al.}(2023)\citenamefont {Wang},
  \citenamefont {Chen}, \citenamefont {Li}, \citenamefont {Qi}, \citenamefont
  {Yu},\ and\ \citenamefont {Guo}}]{wang2023experimental}%
  \BibitemOpen
  \bibfield  {author} {\bibinfo {author} {\bibfnamefont {X.}~\bibnamefont
  {Wang}}, \bibinfo {author} {\bibfnamefont {Z.}~\bibnamefont {Chen}}, \bibinfo
  {author} {\bibfnamefont {Z.}~\bibnamefont {Li}}, \bibinfo {author}
  {\bibfnamefont {D.}~\bibnamefont {Qi}}, \bibinfo {author} {\bibfnamefont
  {S.}~\bibnamefont {Yu}},\ and\ \bibinfo {author} {\bibfnamefont
  {H.}~\bibnamefont {Guo}},\ }\bibfield  {title} {\bibinfo {title}
  {Experimental upstream transmission of continuous variable quantum key
  distribution access network},\ }\href@noop {} {\bibfield  {journal} {\bibinfo
   {journal} {Optics Letters}\ }\textbf {\bibinfo {volume} {48}},\ \bibinfo
  {pages} {3327} (\bibinfo {year} {2023})}\BibitemShut {NoStop}%
\bibitem [{\citenamefont {Proctor}\ \emph {et~al.}(2018)\citenamefont
  {Proctor}, \citenamefont {Knott},\ and\ \citenamefont
  {Dunningham}}]{proctor2018multiparameter}%
  \BibitemOpen
  \bibfield  {author} {\bibinfo {author} {\bibfnamefont {T.~J.}\ \bibnamefont
  {Proctor}}, \bibinfo {author} {\bibfnamefont {P.~A.}\ \bibnamefont {Knott}},\
  and\ \bibinfo {author} {\bibfnamefont {J.~A.}\ \bibnamefont {Dunningham}},\
  }\bibfield  {title} {\bibinfo {title} {Multiparameter estimation in networked
  quantum sensors},\ }\href@noop {} {\bibfield  {journal} {\bibinfo  {journal}
  {Physical review letters}\ }\textbf {\bibinfo {volume} {120}},\ \bibinfo
  {pages} {080501} (\bibinfo {year} {2018})}\BibitemShut {NoStop}%
\bibitem [{\citenamefont {Eldredge}\ \emph {et~al.}(2018)\citenamefont
  {Eldredge}, \citenamefont {Foss-Feig}, \citenamefont {Gross}, \citenamefont
  {Rolston},\ and\ \citenamefont {Gorshkov}}]{eldredge2018optimal}%
  \BibitemOpen
  \bibfield  {author} {\bibinfo {author} {\bibfnamefont {Z.}~\bibnamefont
  {Eldredge}}, \bibinfo {author} {\bibfnamefont {M.}~\bibnamefont {Foss-Feig}},
  \bibinfo {author} {\bibfnamefont {J.~A.}\ \bibnamefont {Gross}}, \bibinfo
  {author} {\bibfnamefont {S.~L.}\ \bibnamefont {Rolston}},\ and\ \bibinfo
  {author} {\bibfnamefont {A.~V.}\ \bibnamefont {Gorshkov}},\ }\bibfield
  {title} {\bibinfo {title} {Optimal and secure measurement protocols for
  quantum sensor networks},\ }\href@noop {} {\bibfield  {journal} {\bibinfo
  {journal} {Physical Review A}\ }\textbf {\bibinfo {volume} {97}},\ \bibinfo
  {pages} {042337} (\bibinfo {year} {2018})}\BibitemShut {NoStop}%
\bibitem [{\citenamefont {Xia}\ \emph {et~al.}(2020)\citenamefont {Xia},
  \citenamefont {Li}, \citenamefont {Clark}, \citenamefont {Hart},
  \citenamefont {Zhuang},\ and\ \citenamefont {Zhang}}]{xia2020demonstration}%
  \BibitemOpen
  \bibfield  {author} {\bibinfo {author} {\bibfnamefont {Y.}~\bibnamefont
  {Xia}}, \bibinfo {author} {\bibfnamefont {W.}~\bibnamefont {Li}}, \bibinfo
  {author} {\bibfnamefont {W.}~\bibnamefont {Clark}}, \bibinfo {author}
  {\bibfnamefont {D.}~\bibnamefont {Hart}}, \bibinfo {author} {\bibfnamefont
  {Q.}~\bibnamefont {Zhuang}},\ and\ \bibinfo {author} {\bibfnamefont
  {Z.}~\bibnamefont {Zhang}},\ }\bibfield  {title} {\bibinfo {title}
  {Demonstration of a reconfigurable entangled radio-frequency photonic sensor
  network},\ }\href@noop {} {\bibfield  {journal} {\bibinfo  {journal}
  {Physical Review Letters}\ }\textbf {\bibinfo {volume} {124}},\ \bibinfo
  {pages} {150502} (\bibinfo {year} {2020})}\BibitemShut {NoStop}%
\bibitem [{\citenamefont {Liu}\ \emph {et~al.}(2021)\citenamefont {Liu},
  \citenamefont {Zhang}, \citenamefont {Li}, \citenamefont {Zhang},
  \citenamefont {Yin}, \citenamefont {Fei}, \citenamefont {Li}, \citenamefont
  {Liu}, \citenamefont {Xu}, \citenamefont {Chen} \emph
  {et~al.}}]{liu2021distributed}%
  \BibitemOpen
  \bibfield  {author} {\bibinfo {author} {\bibfnamefont {L.-Z.}\ \bibnamefont
  {Liu}}, \bibinfo {author} {\bibfnamefont {Y.-Z.}\ \bibnamefont {Zhang}},
  \bibinfo {author} {\bibfnamefont {Z.-D.}\ \bibnamefont {Li}}, \bibinfo
  {author} {\bibfnamefont {R.}~\bibnamefont {Zhang}}, \bibinfo {author}
  {\bibfnamefont {X.-F.}\ \bibnamefont {Yin}}, \bibinfo {author} {\bibfnamefont
  {Y.-Y.}\ \bibnamefont {Fei}}, \bibinfo {author} {\bibfnamefont
  {L.}~\bibnamefont {Li}}, \bibinfo {author} {\bibfnamefont {N.-L.}\
  \bibnamefont {Liu}}, \bibinfo {author} {\bibfnamefont {F.}~\bibnamefont
  {Xu}}, \bibinfo {author} {\bibfnamefont {Y.-A.}\ \bibnamefont {Chen}}, \emph
  {et~al.},\ }\bibfield  {title} {\bibinfo {title} {Distributed quantum phase
  estimation with entangled photons},\ }\href@noop {} {\bibfield  {journal}
  {\bibinfo  {journal} {Nature photonics}\ }\textbf {\bibinfo {volume} {15}},\
  \bibinfo {pages} {137} (\bibinfo {year} {2021})}\BibitemShut {NoStop}%
\bibitem [{\citenamefont {Cirac}\ \emph {et~al.}(1999)\citenamefont {Cirac},
  \citenamefont {Ekert}, \citenamefont {Huelga},\ and\ \citenamefont
  {Macchiavello}}]{cirac1999distributed}%
  \BibitemOpen
  \bibfield  {author} {\bibinfo {author} {\bibfnamefont {J.~I.}\ \bibnamefont
  {Cirac}}, \bibinfo {author} {\bibfnamefont {A.}~\bibnamefont {Ekert}},
  \bibinfo {author} {\bibfnamefont {S.~F.}\ \bibnamefont {Huelga}},\ and\
  \bibinfo {author} {\bibfnamefont {C.}~\bibnamefont {Macchiavello}},\
  }\bibfield  {title} {\bibinfo {title} {Distributed quantum computation over
  noisy channels},\ }\href@noop {} {\bibfield  {journal} {\bibinfo  {journal}
  {Physical Review A}\ }\textbf {\bibinfo {volume} {59}},\ \bibinfo {pages}
  {4249} (\bibinfo {year} {1999})}\BibitemShut {NoStop}%
\bibitem [{\citenamefont {Jiang}\ \emph {et~al.}(2007)\citenamefont {Jiang},
  \citenamefont {Taylor}, \citenamefont {S{\o}rensen},\ and\ \citenamefont
  {Lukin}}]{jiang2007distributed}%
  \BibitemOpen
  \bibfield  {author} {\bibinfo {author} {\bibfnamefont {L.}~\bibnamefont
  {Jiang}}, \bibinfo {author} {\bibfnamefont {J.~M.}\ \bibnamefont {Taylor}},
  \bibinfo {author} {\bibfnamefont {A.~S.}\ \bibnamefont {S{\o}rensen}},\ and\
  \bibinfo {author} {\bibfnamefont {M.~D.}\ \bibnamefont {Lukin}},\ }\bibfield
  {title} {\bibinfo {title} {Distributed quantum computation based on small
  quantum registers},\ }\href@noop {} {\bibfield  {journal} {\bibinfo
  {journal} {Physical Review A}\ }\textbf {\bibinfo {volume} {76}},\ \bibinfo
  {pages} {062323} (\bibinfo {year} {2007})}\BibitemShut {NoStop}%
\bibitem [{\citenamefont {Van~Meter}\ \emph {et~al.}(2010)\citenamefont
  {Van~Meter}, \citenamefont {Ladd}, \citenamefont {Fowler},\ and\
  \citenamefont {Yamamoto}}]{van2010distributed}%
  \BibitemOpen
  \bibfield  {author} {\bibinfo {author} {\bibfnamefont {R.}~\bibnamefont
  {Van~Meter}}, \bibinfo {author} {\bibfnamefont {T.~D.}\ \bibnamefont {Ladd}},
  \bibinfo {author} {\bibfnamefont {A.~G.}\ \bibnamefont {Fowler}},\ and\
  \bibinfo {author} {\bibfnamefont {Y.}~\bibnamefont {Yamamoto}},\ }\bibfield
  {title} {\bibinfo {title} {Distributed quantum computation architecture using
  semiconductor nanophotonics},\ }\href@noop {} {\bibfield  {journal} {\bibinfo
   {journal} {International Journal of Quantum Information}\ }\textbf {\bibinfo
  {volume} {8}},\ \bibinfo {pages} {295} (\bibinfo {year} {2010})}\BibitemShut
  {NoStop}%
\bibitem [{\citenamefont {Monroe}\ \emph {et~al.}(2014)\citenamefont {Monroe},
  \citenamefont {Raussendorf}, \citenamefont {Ruthven}, \citenamefont {Brown},
  \citenamefont {Maunz}, \citenamefont {Duan},\ and\ \citenamefont
  {Kim}}]{monroe2014large}%
  \BibitemOpen
  \bibfield  {author} {\bibinfo {author} {\bibfnamefont {C.}~\bibnamefont
  {Monroe}}, \bibinfo {author} {\bibfnamefont {R.}~\bibnamefont {Raussendorf}},
  \bibinfo {author} {\bibfnamefont {A.}~\bibnamefont {Ruthven}}, \bibinfo
  {author} {\bibfnamefont {K.~R.}\ \bibnamefont {Brown}}, \bibinfo {author}
  {\bibfnamefont {P.}~\bibnamefont {Maunz}}, \bibinfo {author} {\bibfnamefont
  {L.-M.}\ \bibnamefont {Duan}},\ and\ \bibinfo {author} {\bibfnamefont
  {J.}~\bibnamefont {Kim}},\ }\bibfield  {title} {\bibinfo {title} {Large-scale
  modular quantum-computer architecture with atomic memory and photonic
  interconnects},\ }\href@noop {} {\bibfield  {journal} {\bibinfo  {journal}
  {Physical Review A}\ }\textbf {\bibinfo {volume} {89}},\ \bibinfo {pages}
  {022317} (\bibinfo {year} {2014})}\BibitemShut {NoStop}%
\bibitem [{\citenamefont {Bennett}\ and\ \citenamefont
  {Brassard}(2014)}]{bennett2014quantum}%
  \BibitemOpen
  \bibfield  {author} {\bibinfo {author} {\bibfnamefont {C.~H.}\ \bibnamefont
  {Bennett}}\ and\ \bibinfo {author} {\bibfnamefont {G.}~\bibnamefont
  {Brassard}},\ }\bibfield  {title} {\bibinfo {title} {Quantum cryptography:
  Public key distribution and coin tossing},\ }\href@noop {} {\bibfield
  {journal} {\bibinfo  {journal} {Theoretical computer science}\ }\textbf
  {\bibinfo {volume} {560}},\ \bibinfo {pages} {7} (\bibinfo {year}
  {2014})}\BibitemShut {NoStop}%
\bibitem [{\citenamefont {Lucamarini}\ \emph {et~al.}(2018)\citenamefont
  {Lucamarini}, \citenamefont {Yuan}, \citenamefont {Dynes},\ and\
  \citenamefont {Shields}}]{lucamarini2018overcoming}%
  \BibitemOpen
  \bibfield  {author} {\bibinfo {author} {\bibfnamefont {M.}~\bibnamefont
  {Lucamarini}}, \bibinfo {author} {\bibfnamefont {Z.~L.}\ \bibnamefont
  {Yuan}}, \bibinfo {author} {\bibfnamefont {J.~F.}\ \bibnamefont {Dynes}},\
  and\ \bibinfo {author} {\bibfnamefont {A.~J.}\ \bibnamefont {Shields}},\
  }\bibfield  {title} {\bibinfo {title} {Overcoming the rate--distance limit of
  quantum key distribution without quantum repeaters},\ }\href@noop {}
  {\bibfield  {journal} {\bibinfo  {journal} {Nature}\ }\textbf {\bibinfo
  {volume} {557}},\ \bibinfo {pages} {400} (\bibinfo {year}
  {2018})}\BibitemShut {NoStop}%
\bibitem [{\citenamefont {Yin}\ \emph {et~al.}(2020)\citenamefont {Yin},
  \citenamefont {Li}, \citenamefont {Liao}, \citenamefont {Yang}, \citenamefont
  {Cao}, \citenamefont {Zhang}, \citenamefont {Ren}, \citenamefont {Cai},
  \citenamefont {Liu}, \citenamefont {Li} \emph
  {et~al.}}]{yin2020entanglement}%
  \BibitemOpen
  \bibfield  {author} {\bibinfo {author} {\bibfnamefont {J.}~\bibnamefont
  {Yin}}, \bibinfo {author} {\bibfnamefont {Y.-H.}\ \bibnamefont {Li}},
  \bibinfo {author} {\bibfnamefont {S.-K.}\ \bibnamefont {Liao}}, \bibinfo
  {author} {\bibfnamefont {M.}~\bibnamefont {Yang}}, \bibinfo {author}
  {\bibfnamefont {Y.}~\bibnamefont {Cao}}, \bibinfo {author} {\bibfnamefont
  {L.}~\bibnamefont {Zhang}}, \bibinfo {author} {\bibfnamefont {J.-G.}\
  \bibnamefont {Ren}}, \bibinfo {author} {\bibfnamefont {W.-Q.}\ \bibnamefont
  {Cai}}, \bibinfo {author} {\bibfnamefont {W.-Y.}\ \bibnamefont {Liu}},
  \bibinfo {author} {\bibfnamefont {S.-L.}\ \bibnamefont {Li}}, \emph
  {et~al.},\ }\bibfield  {title} {\bibinfo {title} {Entanglement-based secure
  quantum cryptography over 1,120 kilometres},\ }\href@noop {} {\bibfield
  {journal} {\bibinfo  {journal} {Nature}\ }\textbf {\bibinfo {volume} {582}},\
  \bibinfo {pages} {501} (\bibinfo {year} {2020})}\BibitemShut {NoStop}%
\bibitem [{\citenamefont {Wang}\ \emph
  {et~al.}(2022{\natexlab{a}})\citenamefont {Wang}, \citenamefont {Yin},
  \citenamefont {He}, \citenamefont {Chen}, \citenamefont {Wang}, \citenamefont
  {Ye}, \citenamefont {Zhou}, \citenamefont {Fan-Yuan}, \citenamefont {Wang},
  \citenamefont {Chen} \emph {et~al.}}]{wang2022twin}%
  \BibitemOpen
  \bibfield  {author} {\bibinfo {author} {\bibfnamefont {S.}~\bibnamefont
  {Wang}}, \bibinfo {author} {\bibfnamefont {Z.-Q.}\ \bibnamefont {Yin}},
  \bibinfo {author} {\bibfnamefont {D.-Y.}\ \bibnamefont {He}}, \bibinfo
  {author} {\bibfnamefont {W.}~\bibnamefont {Chen}}, \bibinfo {author}
  {\bibfnamefont {R.-Q.}\ \bibnamefont {Wang}}, \bibinfo {author}
  {\bibfnamefont {P.}~\bibnamefont {Ye}}, \bibinfo {author} {\bibfnamefont
  {Y.}~\bibnamefont {Zhou}}, \bibinfo {author} {\bibfnamefont {G.-J.}\
  \bibnamefont {Fan-Yuan}}, \bibinfo {author} {\bibfnamefont {F.-X.}\
  \bibnamefont {Wang}}, \bibinfo {author} {\bibfnamefont {W.}~\bibnamefont
  {Chen}}, \emph {et~al.},\ }\bibfield  {title} {\bibinfo {title} {Twin-field
  quantum key distribution over 830-km fibre},\ }\href@noop {} {\bibfield
  {journal} {\bibinfo  {journal} {Nature photonics}\ }\textbf {\bibinfo
  {volume} {16}},\ \bibinfo {pages} {154} (\bibinfo {year}
  {2022}{\natexlab{a}})}\BibitemShut {NoStop}%
\bibitem [{\citenamefont {Fan-Yuan}\ \emph {et~al.}(2022)\citenamefont
  {Fan-Yuan}, \citenamefont {Lu}, \citenamefont {Wang}, \citenamefont {Yin},
  \citenamefont {He}, \citenamefont {Chen}, \citenamefont {Zhou}, \citenamefont
  {Wang}, \citenamefont {Teng}, \citenamefont {Guo} \emph
  {et~al.}}]{fan2022robust}%
  \BibitemOpen
  \bibfield  {author} {\bibinfo {author} {\bibfnamefont {G.-J.}\ \bibnamefont
  {Fan-Yuan}}, \bibinfo {author} {\bibfnamefont {F.-Y.}\ \bibnamefont {Lu}},
  \bibinfo {author} {\bibfnamefont {S.}~\bibnamefont {Wang}}, \bibinfo {author}
  {\bibfnamefont {Z.-Q.}\ \bibnamefont {Yin}}, \bibinfo {author} {\bibfnamefont
  {D.-Y.}\ \bibnamefont {He}}, \bibinfo {author} {\bibfnamefont
  {W.}~\bibnamefont {Chen}}, \bibinfo {author} {\bibfnamefont {Z.}~\bibnamefont
  {Zhou}}, \bibinfo {author} {\bibfnamefont {Z.-H.}\ \bibnamefont {Wang}},
  \bibinfo {author} {\bibfnamefont {J.}~\bibnamefont {Teng}}, \bibinfo {author}
  {\bibfnamefont {G.-C.}\ \bibnamefont {Guo}}, \emph {et~al.},\ }\bibfield
  {title} {\bibinfo {title} {Robust and adaptable quantum key distribution
  network without trusted nodes},\ }\href@noop {} {\bibfield  {journal}
  {\bibinfo  {journal} {Optica}\ }\textbf {\bibinfo {volume} {9}},\ \bibinfo
  {pages} {812} (\bibinfo {year} {2022})}\BibitemShut {NoStop}%
\bibitem [{\citenamefont {Liu}\ \emph {et~al.}(2023)\citenamefont {Liu},
  \citenamefont {Zhang}, \citenamefont {Jiang}, \citenamefont {Chen},
  \citenamefont {Zhang}, \citenamefont {Pan}, \citenamefont {Ma}, \citenamefont
  {Dong}, \citenamefont {Xiong}, \citenamefont {Zhang} \emph
  {et~al.}}]{liu2023experimental}%
  \BibitemOpen
  \bibfield  {author} {\bibinfo {author} {\bibfnamefont {Y.}~\bibnamefont
  {Liu}}, \bibinfo {author} {\bibfnamefont {W.-J.}\ \bibnamefont {Zhang}},
  \bibinfo {author} {\bibfnamefont {C.}~\bibnamefont {Jiang}}, \bibinfo
  {author} {\bibfnamefont {J.-P.}\ \bibnamefont {Chen}}, \bibinfo {author}
  {\bibfnamefont {C.}~\bibnamefont {Zhang}}, \bibinfo {author} {\bibfnamefont
  {W.-X.}\ \bibnamefont {Pan}}, \bibinfo {author} {\bibfnamefont
  {D.}~\bibnamefont {Ma}}, \bibinfo {author} {\bibfnamefont {H.}~\bibnamefont
  {Dong}}, \bibinfo {author} {\bibfnamefont {J.-M.}\ \bibnamefont {Xiong}},
  \bibinfo {author} {\bibfnamefont {C.-J.}\ \bibnamefont {Zhang}}, \emph
  {et~al.},\ }\bibfield  {title} {\bibinfo {title} {Experimental twin-field
  quantum key distribution over 1000 km fiber distance},\ }\href@noop {}
  {\bibfield  {journal} {\bibinfo  {journal} {Physical Review Letters}\
  }\textbf {\bibinfo {volume} {130}},\ \bibinfo {pages} {210801} (\bibinfo
  {year} {2023})}\BibitemShut {NoStop}%
\bibitem [{\citenamefont {Li}\ \emph {et~al.}(2023)\citenamefont {Li},
  \citenamefont {Zhang}, \citenamefont {Tan}, \citenamefont {Lu}, \citenamefont
  {Liao}, \citenamefont {Huang}, \citenamefont {Li}, \citenamefont {Wang},
  \citenamefont {Mao}, \citenamefont {Yan} \emph {et~al.}}]{li2023high}%
  \BibitemOpen
  \bibfield  {author} {\bibinfo {author} {\bibfnamefont {W.}~\bibnamefont
  {Li}}, \bibinfo {author} {\bibfnamefont {L.}~\bibnamefont {Zhang}}, \bibinfo
  {author} {\bibfnamefont {H.}~\bibnamefont {Tan}}, \bibinfo {author}
  {\bibfnamefont {Y.}~\bibnamefont {Lu}}, \bibinfo {author} {\bibfnamefont
  {S.-K.}\ \bibnamefont {Liao}}, \bibinfo {author} {\bibfnamefont
  {J.}~\bibnamefont {Huang}}, \bibinfo {author} {\bibfnamefont
  {H.}~\bibnamefont {Li}}, \bibinfo {author} {\bibfnamefont {Z.}~\bibnamefont
  {Wang}}, \bibinfo {author} {\bibfnamefont {H.-K.}\ \bibnamefont {Mao}},
  \bibinfo {author} {\bibfnamefont {B.}~\bibnamefont {Yan}}, \emph {et~al.},\
  }\bibfield  {title} {\bibinfo {title} {High-rate quantum key distribution
  exceeding 110 mb s--1},\ }\href@noop {} {\bibfield  {journal} {\bibinfo
  {journal} {Nature photonics}\ }\textbf {\bibinfo {volume} {17}},\ \bibinfo
  {pages} {416} (\bibinfo {year} {2023})}\BibitemShut {NoStop}%
\bibitem [{\citenamefont {Huang}\ \emph {et~al.}(2024)\citenamefont {Huang},
  \citenamefont {Chen}, \citenamefont {Luo}, \citenamefont {He}, \citenamefont
  {Liu}, \citenamefont {Zhang},\ and\ \citenamefont {Wei}}]{huang2024cost}%
  \BibitemOpen
  \bibfield  {author} {\bibinfo {author} {\bibfnamefont {C.}~\bibnamefont
  {Huang}}, \bibinfo {author} {\bibfnamefont {Y.}~\bibnamefont {Chen}},
  \bibinfo {author} {\bibfnamefont {T.}~\bibnamefont {Luo}}, \bibinfo {author}
  {\bibfnamefont {W.}~\bibnamefont {He}}, \bibinfo {author} {\bibfnamefont
  {X.}~\bibnamefont {Liu}}, \bibinfo {author} {\bibfnamefont {Z.}~\bibnamefont
  {Zhang}},\ and\ \bibinfo {author} {\bibfnamefont {K.}~\bibnamefont {Wei}},\
  }\bibfield  {title} {\bibinfo {title} {A cost-efficient quantum access
  network with qubit-based synchronization},\ }\href@noop {} {\bibfield
  {journal} {\bibinfo  {journal} {Science China Physics, Mechanics \&
  Astronomy}\ }\textbf {\bibinfo {volume} {67}},\ \bibinfo {pages} {240312}
  (\bibinfo {year} {2024})}\BibitemShut {NoStop}%
\bibitem [{\citenamefont {Grosshans}\ and\ \citenamefont
  {Grangier}(2002)}]{grosshans2002continuous}%
  \BibitemOpen
  \bibfield  {author} {\bibinfo {author} {\bibfnamefont {F.}~\bibnamefont
  {Grosshans}}\ and\ \bibinfo {author} {\bibfnamefont {P.}~\bibnamefont
  {Grangier}},\ }\bibfield  {title} {\bibinfo {title} {Continuous variable
  quantum cryptography using coherent states},\ }\href@noop {} {\bibfield
  {journal} {\bibinfo  {journal} {Physical review letters}\ }\textbf {\bibinfo
  {volume} {88}},\ \bibinfo {pages} {057902} (\bibinfo {year}
  {2002})}\BibitemShut {NoStop}%
\bibitem [{\citenamefont {Qi}\ \emph {et~al.}(2015)\citenamefont {Qi},
  \citenamefont {Lougovski}, \citenamefont {Pooser}, \citenamefont {Grice},\
  and\ \citenamefont {Bobrek}}]{qi2015generating}%
  \BibitemOpen
  \bibfield  {author} {\bibinfo {author} {\bibfnamefont {B.}~\bibnamefont
  {Qi}}, \bibinfo {author} {\bibfnamefont {P.}~\bibnamefont {Lougovski}},
  \bibinfo {author} {\bibfnamefont {R.}~\bibnamefont {Pooser}}, \bibinfo
  {author} {\bibfnamefont {W.}~\bibnamefont {Grice}},\ and\ \bibinfo {author}
  {\bibfnamefont {M.}~\bibnamefont {Bobrek}},\ }\bibfield  {title} {\bibinfo
  {title} {Generating the local oscillator ''locally'' in continuous-variable
  quantum key distribution based on coherent detection},\ }\href@noop {}
  {\bibfield  {journal} {\bibinfo  {journal} {Physical Review X}\ }\textbf
  {\bibinfo {volume} {5}},\ \bibinfo {pages} {041009} (\bibinfo {year}
  {2015})}\BibitemShut {NoStop}%
\bibitem [{\citenamefont {Huang}\ \emph {et~al.}(2016)\citenamefont {Huang},
  \citenamefont {Huang}, \citenamefont {Lin},\ and\ \citenamefont
  {Zeng}}]{huang2016long}%
  \BibitemOpen
  \bibfield  {author} {\bibinfo {author} {\bibfnamefont {D.}~\bibnamefont
  {Huang}}, \bibinfo {author} {\bibfnamefont {P.}~\bibnamefont {Huang}},
  \bibinfo {author} {\bibfnamefont {D.}~\bibnamefont {Lin}},\ and\ \bibinfo
  {author} {\bibfnamefont {G.}~\bibnamefont {Zeng}},\ }\bibfield  {title}
  {\bibinfo {title} {Long-distance continuous-variable quantum key distribution
  by controlling excess noise},\ }\href@noop {} {\bibfield  {journal} {\bibinfo
   {journal} {Scientific reports}\ }\textbf {\bibinfo {volume} {6}},\ \bibinfo
  {pages} {19201} (\bibinfo {year} {2016})}\BibitemShut {NoStop}%
\bibitem [{\citenamefont {Zhang}\ \emph {et~al.}(2020)\citenamefont {Zhang},
  \citenamefont {Chen}, \citenamefont {Pirandola}, \citenamefont {Wang},
  \citenamefont {Zhou}, \citenamefont {Chu}, \citenamefont {Zhao},
  \citenamefont {Xu}, \citenamefont {Yu},\ and\ \citenamefont
  {Guo}}]{zhang2020long}%
  \BibitemOpen
  \bibfield  {author} {\bibinfo {author} {\bibfnamefont {Y.}~\bibnamefont
  {Zhang}}, \bibinfo {author} {\bibfnamefont {Z.}~\bibnamefont {Chen}},
  \bibinfo {author} {\bibfnamefont {S.}~\bibnamefont {Pirandola}}, \bibinfo
  {author} {\bibfnamefont {X.}~\bibnamefont {Wang}}, \bibinfo {author}
  {\bibfnamefont {C.}~\bibnamefont {Zhou}}, \bibinfo {author} {\bibfnamefont
  {B.}~\bibnamefont {Chu}}, \bibinfo {author} {\bibfnamefont {Y.}~\bibnamefont
  {Zhao}}, \bibinfo {author} {\bibfnamefont {B.}~\bibnamefont {Xu}}, \bibinfo
  {author} {\bibfnamefont {S.}~\bibnamefont {Yu}},\ and\ \bibinfo {author}
  {\bibfnamefont {H.}~\bibnamefont {Guo}},\ }\bibfield  {title} {\bibinfo
  {title} {Long-distance continuous-variable quantum key distribution over
  202.81 km of fiber},\ }\href@noop {} {\bibfield  {journal} {\bibinfo
  {journal} {Physical review letters}\ }\textbf {\bibinfo {volume} {125}},\
  \bibinfo {pages} {010502} (\bibinfo {year} {2020})}\BibitemShut {NoStop}%
\bibitem [{\citenamefont {Wang}\ \emph
  {et~al.}(2022{\natexlab{b}})\citenamefont {Wang}, \citenamefont {Li},
  \citenamefont {Pi}, \citenamefont {Pan}, \citenamefont {Shao}, \citenamefont
  {Ma}, \citenamefont {Zhang}, \citenamefont {Yang}, \citenamefont {Zhang},
  \citenamefont {Huang} \emph {et~al.}}]{wang2022sub}%
  \BibitemOpen
  \bibfield  {author} {\bibinfo {author} {\bibfnamefont {H.}~\bibnamefont
  {Wang}}, \bibinfo {author} {\bibfnamefont {Y.}~\bibnamefont {Li}}, \bibinfo
  {author} {\bibfnamefont {Y.}~\bibnamefont {Pi}}, \bibinfo {author}
  {\bibfnamefont {Y.}~\bibnamefont {Pan}}, \bibinfo {author} {\bibfnamefont
  {Y.}~\bibnamefont {Shao}}, \bibinfo {author} {\bibfnamefont {L.}~\bibnamefont
  {Ma}}, \bibinfo {author} {\bibfnamefont {Y.}~\bibnamefont {Zhang}}, \bibinfo
  {author} {\bibfnamefont {J.}~\bibnamefont {Yang}}, \bibinfo {author}
  {\bibfnamefont {T.}~\bibnamefont {Zhang}}, \bibinfo {author} {\bibfnamefont
  {W.}~\bibnamefont {Huang}}, \emph {et~al.},\ }\bibfield  {title} {\bibinfo
  {title} {Sub-gbps key rate four-state continuous-variable quantum key
  distribution within metropolitan area},\ }\href@noop {} {\bibfield  {journal}
  {\bibinfo  {journal} {Communications Physics}\ }\textbf {\bibinfo {volume}
  {5}},\ \bibinfo {pages} {162} (\bibinfo {year}
  {2022}{\natexlab{b}})}\BibitemShut {NoStop}%
\bibitem [{\citenamefont {Xu}\ \emph {et~al.}(2023)\citenamefont {Xu},
  \citenamefont {Wang}, \citenamefont {Zhao}, \citenamefont {Huang},\ and\
  \citenamefont {Zeng}}]{xu2023round}%
  \BibitemOpen
  \bibfield  {author} {\bibinfo {author} {\bibfnamefont {Y.}~\bibnamefont
  {Xu}}, \bibinfo {author} {\bibfnamefont {T.}~\bibnamefont {Wang}}, \bibinfo
  {author} {\bibfnamefont {H.}~\bibnamefont {Zhao}}, \bibinfo {author}
  {\bibfnamefont {P.}~\bibnamefont {Huang}},\ and\ \bibinfo {author}
  {\bibfnamefont {G.}~\bibnamefont {Zeng}},\ }\bibfield  {title} {\bibinfo
  {title} {Round-trip multi-band quantum access network},\ }\href@noop {}
  {\bibfield  {journal} {\bibinfo  {journal} {Photonics Research}\ }\textbf
  {\bibinfo {volume} {11}},\ \bibinfo {pages} {1449} (\bibinfo {year}
  {2023})}\BibitemShut {NoStop}%
\bibitem [{\citenamefont {Hajomer}\ \emph
  {et~al.}(2024{\natexlab{a}})\citenamefont {Hajomer}, \citenamefont {Derkach},
  \citenamefont {Filip}, \citenamefont {Andersen}, \citenamefont {C.~Usenko},\
  and\ \citenamefont {Gehring}}]{hajomer2024continuous}%
  \BibitemOpen
  \bibfield  {author} {\bibinfo {author} {\bibfnamefont {A.~A.}\ \bibnamefont
  {Hajomer}}, \bibinfo {author} {\bibfnamefont {I.}~\bibnamefont {Derkach}},
  \bibinfo {author} {\bibfnamefont {R.}~\bibnamefont {Filip}}, \bibinfo
  {author} {\bibfnamefont {U.~L.}\ \bibnamefont {Andersen}}, \bibinfo {author}
  {\bibfnamefont {V.}~\bibnamefont {C.~Usenko}},\ and\ \bibinfo {author}
  {\bibfnamefont {T.}~\bibnamefont {Gehring}},\ }\bibfield  {title} {\bibinfo
  {title} {Continuous-variable quantum passive optical network},\ }\href@noop
  {} {\bibfield  {journal} {\bibinfo  {journal} {Light: Science \&
  Applications}\ }\textbf {\bibinfo {volume} {13}},\ \bibinfo {pages} {291}
  (\bibinfo {year} {2024}{\natexlab{a}})}\BibitemShut {NoStop}%
\bibitem [{\citenamefont {Liu}\ \emph {et~al.}(2024{\natexlab{a}})\citenamefont
  {Liu}, \citenamefont {Tian}, \citenamefont {Zhang}, \citenamefont {Lu},
  \citenamefont {Wang},\ and\ \citenamefont {Li}}]{liu2024integrated}%
  \BibitemOpen
  \bibfield  {author} {\bibinfo {author} {\bibfnamefont {S.}~\bibnamefont
  {Liu}}, \bibinfo {author} {\bibfnamefont {Y.}~\bibnamefont {Tian}}, \bibinfo
  {author} {\bibfnamefont {Y.}~\bibnamefont {Zhang}}, \bibinfo {author}
  {\bibfnamefont {Z.}~\bibnamefont {Lu}}, \bibinfo {author} {\bibfnamefont
  {X.}~\bibnamefont {Wang}},\ and\ \bibinfo {author} {\bibfnamefont
  {Y.}~\bibnamefont {Li}},\ }\bibfield  {title} {\bibinfo {title} {Integrated
  quantum communication network and vibration sensing in optical fibers},\
  }\href@noop {} {\bibfield  {journal} {\bibinfo  {journal} {Optica}\ }\textbf
  {\bibinfo {volume} {11}},\ \bibinfo {pages} {1762} (\bibinfo {year}
  {2024}{\natexlab{a}})}\BibitemShut {NoStop}%
\bibitem [{\citenamefont {Hajomer}\ \emph
  {et~al.}(2024{\natexlab{b}})\citenamefont {Hajomer}, \citenamefont {Derkach},
  \citenamefont {Jain}, \citenamefont {Chin}, \citenamefont {Andersen},\ and\
  \citenamefont {Gehring}}]{hajomer2024long}%
  \BibitemOpen
  \bibfield  {author} {\bibinfo {author} {\bibfnamefont {A.~A.}\ \bibnamefont
  {Hajomer}}, \bibinfo {author} {\bibfnamefont {I.}~\bibnamefont {Derkach}},
  \bibinfo {author} {\bibfnamefont {N.}~\bibnamefont {Jain}}, \bibinfo {author}
  {\bibfnamefont {H.-M.}\ \bibnamefont {Chin}}, \bibinfo {author}
  {\bibfnamefont {U.~L.}\ \bibnamefont {Andersen}},\ and\ \bibinfo {author}
  {\bibfnamefont {T.}~\bibnamefont {Gehring}},\ }\bibfield  {title} {\bibinfo
  {title} {Long-distance continuous-variable quantum key distribution over
  100-km fiber with local local oscillator},\ }\href@noop {} {\bibfield
  {journal} {\bibinfo  {journal} {Science Advances}\ }\textbf {\bibinfo
  {volume} {10}},\ \bibinfo {pages} {eadi9474} (\bibinfo {year}
  {2024}{\natexlab{b}})}\BibitemShut {NoStop}%
\bibitem [{\citenamefont {Xu}\ \emph {et~al.}(2024{\natexlab{a}})\citenamefont
  {Xu}, \citenamefont {Wang}, \citenamefont {Huang},\ and\ \citenamefont
  {Zeng}}]{research.0416}%
  \BibitemOpen
  \bibfield  {author} {\bibinfo {author} {\bibfnamefont {Y.}~\bibnamefont
  {Xu}}, \bibinfo {author} {\bibfnamefont {T.}~\bibnamefont {Wang}}, \bibinfo
  {author} {\bibfnamefont {P.}~\bibnamefont {Huang}},\ and\ \bibinfo {author}
  {\bibfnamefont {G.}~\bibnamefont {Zeng}},\ }\bibfield  {title} {\bibinfo
  {title} {Integrated distributed sensing and quantum communication networks},\
  }\href@noop {} {\bibfield  {journal} {\bibinfo  {journal} {Research}\
  }\textbf {\bibinfo {volume} {7}},\ \bibinfo {pages} {0416} (\bibinfo {year}
  {2024}{\natexlab{a}})}\BibitemShut {NoStop}%
\bibitem [{\citenamefont {Qi}\ \emph {et~al.}(2024)\citenamefont {Qi},
  \citenamefont {Wang}, \citenamefont {Li}, \citenamefont {Ma}, \citenamefont
  {Chen}, \citenamefont {Lu},\ and\ \citenamefont {Yu}}]{qi2024experimental}%
  \BibitemOpen
  \bibfield  {author} {\bibinfo {author} {\bibfnamefont {D.}~\bibnamefont
  {Qi}}, \bibinfo {author} {\bibfnamefont {X.}~\bibnamefont {Wang}}, \bibinfo
  {author} {\bibfnamefont {Z.}~\bibnamefont {Li}}, \bibinfo {author}
  {\bibfnamefont {J.}~\bibnamefont {Ma}}, \bibinfo {author} {\bibfnamefont
  {Z.}~\bibnamefont {Chen}}, \bibinfo {author} {\bibfnamefont {Y.}~\bibnamefont
  {Lu}},\ and\ \bibinfo {author} {\bibfnamefont {S.}~\bibnamefont {Yu}},\
  }\bibfield  {title} {\bibinfo {title} {Experimental demonstration of a
  quantum downstream access network in continuous variable quantum key
  distribution with a local local oscillator},\ }\href@noop {} {\bibfield
  {journal} {\bibinfo  {journal} {Photonics Research}\ }\textbf {\bibinfo
  {volume} {12}},\ \bibinfo {pages} {1262} (\bibinfo {year}
  {2024})}\BibitemShut {NoStop}%
\bibitem [{\citenamefont {Pan}\ \emph {et~al.}(2024)\citenamefont {Pan},
  \citenamefont {Bian}, \citenamefont {Li}, \citenamefont {Xu}, \citenamefont
  {Ma}, \citenamefont {Wang}, \citenamefont {Luo}, \citenamefont {Dou},
  \citenamefont {Pi}, \citenamefont {Yang} \emph {et~al.}}]{pan2024high}%
  \BibitemOpen
  \bibfield  {author} {\bibinfo {author} {\bibfnamefont {Y.}~\bibnamefont
  {Pan}}, \bibinfo {author} {\bibfnamefont {Y.}~\bibnamefont {Bian}}, \bibinfo
  {author} {\bibfnamefont {Y.}~\bibnamefont {Li}}, \bibinfo {author}
  {\bibfnamefont {X.}~\bibnamefont {Xu}}, \bibinfo {author} {\bibfnamefont
  {L.}~\bibnamefont {Ma}}, \bibinfo {author} {\bibfnamefont {H.}~\bibnamefont
  {Wang}}, \bibinfo {author} {\bibfnamefont {Y.}~\bibnamefont {Luo}}, \bibinfo
  {author} {\bibfnamefont {J.}~\bibnamefont {Dou}}, \bibinfo {author}
  {\bibfnamefont {Y.}~\bibnamefont {Pi}}, \bibinfo {author} {\bibfnamefont
  {J.}~\bibnamefont {Yang}}, \emph {et~al.},\ }\bibfield  {title} {\bibinfo
  {title} {High-rate 16-node quantum access network based on passive optical
  network},\ }\href@noop {} {\bibfield  {journal} {\bibinfo  {journal} {arXiv
  preprint arXiv:2403.02585}\ } (\bibinfo {year} {2024})}\BibitemShut {NoStop}%
\bibitem [{\citenamefont {Li}\ \emph {et~al.}(2024)\citenamefont {Li},
  \citenamefont {Wang}, \citenamefont {Qi}, \citenamefont {Chen},\ and\
  \citenamefont {Yu}}]{li2024experimental}%
  \BibitemOpen
  \bibfield  {author} {\bibinfo {author} {\bibfnamefont {Z.}~\bibnamefont
  {Li}}, \bibinfo {author} {\bibfnamefont {X.}~\bibnamefont {Wang}}, \bibinfo
  {author} {\bibfnamefont {D.}~\bibnamefont {Qi}}, \bibinfo {author}
  {\bibfnamefont {Z.}~\bibnamefont {Chen}},\ and\ \bibinfo {author}
  {\bibfnamefont {S.}~\bibnamefont {Yu}},\ }\bibfield  {title} {\bibinfo
  {title} {Experimental implementation of four-user downstream access network
  continuous-variable quantum key distribution},\ }\href@noop {} {\bibfield
  {journal} {\bibinfo  {journal} {Journal of Lightwave Technology}\ } (\bibinfo
  {year} {2024})}\BibitemShut {NoStop}%
\bibitem [{\citenamefont {Brackett}(1990)}]{brackett1990dense}%
  \BibitemOpen
  \bibfield  {author} {\bibinfo {author} {\bibfnamefont {C.~A.}\ \bibnamefont
  {Brackett}},\ }\bibfield  {title} {\bibinfo {title} {Dense wavelength
  division multiplexing networks: Principles and applications},\ }\href@noop {}
  {\bibfield  {journal} {\bibinfo  {journal} {IEEE Journal on Selected areas in
  Communications}\ }\textbf {\bibinfo {volume} {8}},\ \bibinfo {pages} {948}
  (\bibinfo {year} {1990})}\BibitemShut {NoStop}%
\bibitem [{\citenamefont {Banerjee}\ \emph {et~al.}(2005)\citenamefont
  {Banerjee}, \citenamefont {Park}, \citenamefont {Clarke}, \citenamefont
  {Song}, \citenamefont {Yang}, \citenamefont {Kramer}, \citenamefont {Kim},\
  and\ \citenamefont {Mukherjee}}]{banerjee2005wavelength}%
  \BibitemOpen
  \bibfield  {author} {\bibinfo {author} {\bibfnamefont {A.}~\bibnamefont
  {Banerjee}}, \bibinfo {author} {\bibfnamefont {Y.}~\bibnamefont {Park}},
  \bibinfo {author} {\bibfnamefont {F.}~\bibnamefont {Clarke}}, \bibinfo
  {author} {\bibfnamefont {H.}~\bibnamefont {Song}}, \bibinfo {author}
  {\bibfnamefont {S.}~\bibnamefont {Yang}}, \bibinfo {author} {\bibfnamefont
  {G.}~\bibnamefont {Kramer}}, \bibinfo {author} {\bibfnamefont
  {K.}~\bibnamefont {Kim}},\ and\ \bibinfo {author} {\bibfnamefont
  {B.}~\bibnamefont {Mukherjee}},\ }\bibfield  {title} {\bibinfo {title}
  {Wavelength-division-multiplexed passive optical network (wdm-pon)
  technologies for broadband access: a review},\ }\href@noop {} {\bibfield
  {journal} {\bibinfo  {journal} {Journal of optical networking}\ }\textbf
  {\bibinfo {volume} {4}},\ \bibinfo {pages} {737} (\bibinfo {year}
  {2005})}\BibitemShut {NoStop}%
\bibitem [{\citenamefont {Winzer}(2015)}]{winzer2015scaling}%
  \BibitemOpen
  \bibfield  {author} {\bibinfo {author} {\bibfnamefont {P.~J.}\ \bibnamefont
  {Winzer}},\ }\bibfield  {title} {\bibinfo {title} {Scaling optical fiber
  networks: Challenges and solutions},\ }\href@noop {} {\bibfield  {journal}
  {\bibinfo  {journal} {Optics and Photonics News}\ }\textbf {\bibinfo {volume}
  {26}},\ \bibinfo {pages} {28} (\bibinfo {year} {2015})}\BibitemShut {NoStop}%
\bibitem [{\citenamefont {Menicucci}\ \emph {et~al.}(2008)\citenamefont
  {Menicucci}, \citenamefont {Flammia},\ and\ \citenamefont
  {Pfister}}]{menicucci2008one}%
  \BibitemOpen
  \bibfield  {author} {\bibinfo {author} {\bibfnamefont {N.~C.}\ \bibnamefont
  {Menicucci}}, \bibinfo {author} {\bibfnamefont {S.~T.}\ \bibnamefont
  {Flammia}},\ and\ \bibinfo {author} {\bibfnamefont {O.}~\bibnamefont
  {Pfister}},\ }\bibfield  {title} {\bibinfo {title} {One-way quantum computing
  in the optical frequency comb},\ }\href@noop {} {\bibfield  {journal}
  {\bibinfo  {journal} {Physical review letters}\ }\textbf {\bibinfo {volume}
  {101}},\ \bibinfo {pages} {130501} (\bibinfo {year} {2008})}\BibitemShut
  {NoStop}%
\bibitem [{\citenamefont {Foltynowicz}\ \emph {et~al.}(2011)\citenamefont
  {Foltynowicz}, \citenamefont {Ban}, \citenamefont {Mas{\l}owski},
  \citenamefont {Adler},\ and\ \citenamefont {Ye}}]{foltynowicz2011quantum}%
  \BibitemOpen
  \bibfield  {author} {\bibinfo {author} {\bibfnamefont {A.}~\bibnamefont
  {Foltynowicz}}, \bibinfo {author} {\bibfnamefont {T.}~\bibnamefont {Ban}},
  \bibinfo {author} {\bibfnamefont {P.}~\bibnamefont {Mas{\l}owski}}, \bibinfo
  {author} {\bibfnamefont {F.}~\bibnamefont {Adler}},\ and\ \bibinfo {author}
  {\bibfnamefont {J.}~\bibnamefont {Ye}},\ }\bibfield  {title} {\bibinfo
  {title} {Quantum-noise-limited optical frequency comb spectroscopy},\
  }\href@noop {} {\bibfield  {journal} {\bibinfo  {journal} {Physical review
  letters}\ }\textbf {\bibinfo {volume} {107}},\ \bibinfo {pages} {233002}
  (\bibinfo {year} {2011})}\BibitemShut {NoStop}%
\bibitem [{\citenamefont {Roslund}\ \emph {et~al.}(2014)\citenamefont
  {Roslund}, \citenamefont {De~Araujo}, \citenamefont {Jiang}, \citenamefont
  {Fabre},\ and\ \citenamefont {Treps}}]{roslund2014wavelength}%
  \BibitemOpen
  \bibfield  {author} {\bibinfo {author} {\bibfnamefont {J.}~\bibnamefont
  {Roslund}}, \bibinfo {author} {\bibfnamefont {R.~M.}\ \bibnamefont
  {De~Araujo}}, \bibinfo {author} {\bibfnamefont {S.}~\bibnamefont {Jiang}},
  \bibinfo {author} {\bibfnamefont {C.}~\bibnamefont {Fabre}},\ and\ \bibinfo
  {author} {\bibfnamefont {N.}~\bibnamefont {Treps}},\ }\bibfield  {title}
  {\bibinfo {title} {Wavelength-multiplexed quantum networks with ultrafast
  frequency combs},\ }\href@noop {} {\bibfield  {journal} {\bibinfo  {journal}
  {Nature Photonics}\ }\textbf {\bibinfo {volume} {8}},\ \bibinfo {pages} {109}
  (\bibinfo {year} {2014})}\BibitemShut {NoStop}%
\bibitem [{\citenamefont {Pfister}(2019)}]{pfister2019continuous}%
  \BibitemOpen
  \bibfield  {author} {\bibinfo {author} {\bibfnamefont {O.}~\bibnamefont
  {Pfister}},\ }\bibfield  {title} {\bibinfo {title} {Continuous-variable
  quantum computing in the quantum optical frequency comb},\ }\href@noop {}
  {\bibfield  {journal} {\bibinfo  {journal} {Journal of Physics B: Atomic,
  Molecular and Optical Physics}\ }\textbf {\bibinfo {volume} {53}},\ \bibinfo
  {pages} {012001} (\bibinfo {year} {2019})}\BibitemShut {NoStop}%
\bibitem [{\citenamefont {Cai}\ \emph {et~al.}(2021)\citenamefont {Cai},
  \citenamefont {Roslund}, \citenamefont {Thiel}, \citenamefont {Fabre},\ and\
  \citenamefont {Treps}}]{cai2021quantum}%
  \BibitemOpen
  \bibfield  {author} {\bibinfo {author} {\bibfnamefont {Y.}~\bibnamefont
  {Cai}}, \bibinfo {author} {\bibfnamefont {J.}~\bibnamefont {Roslund}},
  \bibinfo {author} {\bibfnamefont {V.}~\bibnamefont {Thiel}}, \bibinfo
  {author} {\bibfnamefont {C.}~\bibnamefont {Fabre}},\ and\ \bibinfo {author}
  {\bibfnamefont {N.}~\bibnamefont {Treps}},\ }\bibfield  {title} {\bibinfo
  {title} {Quantum enhanced measurement of an optical frequency comb},\
  }\href@noop {} {\bibfield  {journal} {\bibinfo  {journal} {npj Quantum
  Information}\ }\textbf {\bibinfo {volume} {7}},\ \bibinfo {pages} {82}
  (\bibinfo {year} {2021})}\BibitemShut {NoStop}%
\bibitem [{\citenamefont {Caldwell}\ \emph {et~al.}(2022)\citenamefont
  {Caldwell}, \citenamefont {Sinclair}, \citenamefont {Newbury},\ and\
  \citenamefont {Deschenes}}]{caldwell2022time}%
  \BibitemOpen
  \bibfield  {author} {\bibinfo {author} {\bibfnamefont {E.~D.}\ \bibnamefont
  {Caldwell}}, \bibinfo {author} {\bibfnamefont {L.~C.}\ \bibnamefont
  {Sinclair}}, \bibinfo {author} {\bibfnamefont {N.~R.}\ \bibnamefont
  {Newbury}},\ and\ \bibinfo {author} {\bibfnamefont {J.-D.}\ \bibnamefont
  {Deschenes}},\ }\bibfield  {title} {\bibinfo {title} {The time-programmable
  frequency comb and its use in quantum-limited ranging},\ }\href@noop {}
  {\bibfield  {journal} {\bibinfo  {journal} {Nature}\ }\textbf {\bibinfo
  {volume} {610}},\ \bibinfo {pages} {667} (\bibinfo {year}
  {2022})}\BibitemShut {NoStop}%
\bibitem [{\citenamefont {Liu}\ \emph {et~al.}(2024{\natexlab{b}})\citenamefont
  {Liu}, \citenamefont {Luo}, \citenamefont {Yu}, \citenamefont {Wang},
  \citenamefont {Wang}, \citenamefont {Hu}, \citenamefont {Li}, \citenamefont
  {Zheng}, \citenamefont {Yao}, \citenamefont {Yan} \emph
  {et~al.}}]{liu2024creation}%
  \BibitemOpen
  \bibfield  {author} {\bibinfo {author} {\bibfnamefont {J.-L.}\ \bibnamefont
  {Liu}}, \bibinfo {author} {\bibfnamefont {X.-Y.}\ \bibnamefont {Luo}},
  \bibinfo {author} {\bibfnamefont {Y.}~\bibnamefont {Yu}}, \bibinfo {author}
  {\bibfnamefont {C.-Y.}\ \bibnamefont {Wang}}, \bibinfo {author}
  {\bibfnamefont {B.}~\bibnamefont {Wang}}, \bibinfo {author} {\bibfnamefont
  {Y.}~\bibnamefont {Hu}}, \bibinfo {author} {\bibfnamefont {J.}~\bibnamefont
  {Li}}, \bibinfo {author} {\bibfnamefont {M.-Y.}\ \bibnamefont {Zheng}},
  \bibinfo {author} {\bibfnamefont {B.}~\bibnamefont {Yao}}, \bibinfo {author}
  {\bibfnamefont {Z.}~\bibnamefont {Yan}}, \emph {et~al.},\ }\bibfield  {title}
  {\bibinfo {title} {Creation of memory--memory entanglement in a metropolitan
  quantum network},\ }\href@noop {} {\bibfield  {journal} {\bibinfo  {journal}
  {Nature}\ }\textbf {\bibinfo {volume} {629}},\ \bibinfo {pages} {579}
  (\bibinfo {year} {2024}{\natexlab{b}})}\BibitemShut {NoStop}%
\bibitem [{\citenamefont {Wang}\ \emph {et~al.}(2025)\citenamefont {Wang},
  \citenamefont {Li}, \citenamefont {Wang}, \citenamefont {Zhou}, \citenamefont
  {Cheng}, \citenamefont {Jing}, \citenamefont {Sun}, \citenamefont {Li},
  \citenamefont {Li}, \citenamefont {Wu} \emph {et~al.}}]{wang2025large}%
  \BibitemOpen
  \bibfield  {author} {\bibinfo {author} {\bibfnamefont {Z.}~\bibnamefont
  {Wang}}, \bibinfo {author} {\bibfnamefont {K.}~\bibnamefont {Li}}, \bibinfo
  {author} {\bibfnamefont {Y.}~\bibnamefont {Wang}}, \bibinfo {author}
  {\bibfnamefont {X.}~\bibnamefont {Zhou}}, \bibinfo {author} {\bibfnamefont
  {Y.}~\bibnamefont {Cheng}}, \bibinfo {author} {\bibfnamefont
  {B.}~\bibnamefont {Jing}}, \bibinfo {author} {\bibfnamefont {F.}~\bibnamefont
  {Sun}}, \bibinfo {author} {\bibfnamefont {J.}~\bibnamefont {Li}}, \bibinfo
  {author} {\bibfnamefont {Z.}~\bibnamefont {Li}}, \bibinfo {author}
  {\bibfnamefont {B.}~\bibnamefont {Wu}}, \emph {et~al.},\ }\bibfield  {title}
  {\bibinfo {title} {Large-scale cluster quantum microcombs},\ }\href@noop {}
  {\bibfield  {journal} {\bibinfo  {journal} {Light: Science \& Applications}\
  }\textbf {\bibinfo {volume} {14}},\ \bibinfo {pages} {164} (\bibinfo {year}
  {2025})}\BibitemShut {NoStop}%
\bibitem [{\citenamefont {Jia}\ \emph {et~al.}(2025)\citenamefont {Jia},
  \citenamefont {Zhai}, \citenamefont {Zhu}, \citenamefont {You}, \citenamefont
  {Cao}, \citenamefont {Zhang}, \citenamefont {Zheng}, \citenamefont {Fu},
  \citenamefont {Mao}, \citenamefont {Dai} \emph {et~al.}}]{jia2025continuous}%
  \BibitemOpen
  \bibfield  {author} {\bibinfo {author} {\bibfnamefont {X.}~\bibnamefont
  {Jia}}, \bibinfo {author} {\bibfnamefont {C.}~\bibnamefont {Zhai}}, \bibinfo
  {author} {\bibfnamefont {X.}~\bibnamefont {Zhu}}, \bibinfo {author}
  {\bibfnamefont {C.}~\bibnamefont {You}}, \bibinfo {author} {\bibfnamefont
  {Y.}~\bibnamefont {Cao}}, \bibinfo {author} {\bibfnamefont {X.}~\bibnamefont
  {Zhang}}, \bibinfo {author} {\bibfnamefont {Y.}~\bibnamefont {Zheng}},
  \bibinfo {author} {\bibfnamefont {Z.}~\bibnamefont {Fu}}, \bibinfo {author}
  {\bibfnamefont {J.}~\bibnamefont {Mao}}, \bibinfo {author} {\bibfnamefont
  {T.}~\bibnamefont {Dai}}, \emph {et~al.},\ }\bibfield  {title} {\bibinfo
  {title} {Continuous-variable multipartite entanglement in an integrated
  microcomb},\ }\href@noop {} {\bibfield  {journal} {\bibinfo  {journal}
  {Nature}\ ,\ \bibinfo {pages} {1}} (\bibinfo {year} {2025})}\BibitemShut
  {NoStop}%
\bibitem [{\citenamefont {Fan}\ \emph {et~al.}(2025)\citenamefont {Fan},
  \citenamefont {Luo}, \citenamefont {Guo}, \citenamefont {Wu}, \citenamefont
  {Zeng}, \citenamefont {Deng}, \citenamefont {Wang}, \citenamefont {Song},
  \citenamefont {Wang}, \citenamefont {You} \emph {et~al.}}]{fan2025quantum}%
  \BibitemOpen
  \bibfield  {author} {\bibinfo {author} {\bibfnamefont {Y.-R.}\ \bibnamefont
  {Fan}}, \bibinfo {author} {\bibfnamefont {Y.}~\bibnamefont {Luo}}, \bibinfo
  {author} {\bibfnamefont {K.}~\bibnamefont {Guo}}, \bibinfo {author}
  {\bibfnamefont {J.-P.}\ \bibnamefont {Wu}}, \bibinfo {author} {\bibfnamefont
  {H.}~\bibnamefont {Zeng}}, \bibinfo {author} {\bibfnamefont {G.-W.}\
  \bibnamefont {Deng}}, \bibinfo {author} {\bibfnamefont {Y.}~\bibnamefont
  {Wang}}, \bibinfo {author} {\bibfnamefont {H.-Z.}\ \bibnamefont {Song}},
  \bibinfo {author} {\bibfnamefont {Z.}~\bibnamefont {Wang}}, \bibinfo {author}
  {\bibfnamefont {L.-X.}\ \bibnamefont {You}}, \emph {et~al.},\ }\bibfield
  {title} {\bibinfo {title} {Quantum entanglement network enabled by a
  state-multiplexing quantum light source},\ }\href@noop {} {\bibfield
  {journal} {\bibinfo  {journal} {Light: Science \& Applications}\ }\textbf
  {\bibinfo {volume} {14}},\ \bibinfo {pages} {1} (\bibinfo {year}
  {2025})}\BibitemShut {NoStop}%
\bibitem [{\citenamefont {Fabre}\ and\ \citenamefont
  {Treps}(2020)}]{fabre2020modes}%
  \BibitemOpen
  \bibfield  {author} {\bibinfo {author} {\bibfnamefont {C.}~\bibnamefont
  {Fabre}}\ and\ \bibinfo {author} {\bibfnamefont {N.}~\bibnamefont {Treps}},\
  }\bibfield  {title} {\bibinfo {title} {Modes and states in quantum optics},\
  }\href@noop {} {\bibfield  {journal} {\bibinfo  {journal} {Reviews of Modern
  Physics}\ }\textbf {\bibinfo {volume} {92}},\ \bibinfo {pages} {035005}
  (\bibinfo {year} {2020})}\BibitemShut {NoStop}%
\bibitem [{\citenamefont {Blow}\ \emph {et~al.}(1990)\citenamefont {Blow},
  \citenamefont {Loudon}, \citenamefont {Phoenix},\ and\ \citenamefont
  {Shepherd}}]{blow1990continuum}%
  \BibitemOpen
  \bibfield  {author} {\bibinfo {author} {\bibfnamefont {K.}~\bibnamefont
  {Blow}}, \bibinfo {author} {\bibfnamefont {R.}~\bibnamefont {Loudon}},
  \bibinfo {author} {\bibfnamefont {S.~J.}\ \bibnamefont {Phoenix}},\ and\
  \bibinfo {author} {\bibfnamefont {T.}~\bibnamefont {Shepherd}},\ }\bibfield
  {title} {\bibinfo {title} {Continuum fields in quantum optics},\ }\href@noop
  {} {\bibfield  {journal} {\bibinfo  {journal} {Physical Review A}\ }\textbf
  {\bibinfo {volume} {42}},\ \bibinfo {pages} {4102} (\bibinfo {year}
  {1990})}\BibitemShut {NoStop}%
\bibitem [{\citenamefont {Raymer}\ and\ \citenamefont
  {Walmsley}(2020)}]{raymer2020temporal}%
  \BibitemOpen
  \bibfield  {author} {\bibinfo {author} {\bibfnamefont {M.~G.}\ \bibnamefont
  {Raymer}}\ and\ \bibinfo {author} {\bibfnamefont {I.~A.}\ \bibnamefont
  {Walmsley}},\ }\bibfield  {title} {\bibinfo {title} {Temporal modes in
  quantum optics: then and now},\ }\href@noop {} {\bibfield  {journal}
  {\bibinfo  {journal} {Physica Scripta}\ }\textbf {\bibinfo {volume} {95}},\
  \bibinfo {pages} {064002} (\bibinfo {year} {2020})}\BibitemShut {NoStop}%
\bibitem [{\citenamefont {Scarani}\ \emph {et~al.}(2009)\citenamefont
  {Scarani}, \citenamefont {Bechmann-Pasquinucci}, \citenamefont {Cerf},
  \citenamefont {Du{\v{s}}ek}, \citenamefont {L{\"u}tkenhaus},\ and\
  \citenamefont {Peev}}]{scarani2009security}%
  \BibitemOpen
  \bibfield  {author} {\bibinfo {author} {\bibfnamefont {V.}~\bibnamefont
  {Scarani}}, \bibinfo {author} {\bibfnamefont {H.}~\bibnamefont
  {Bechmann-Pasquinucci}}, \bibinfo {author} {\bibfnamefont {N.~J.}\
  \bibnamefont {Cerf}}, \bibinfo {author} {\bibfnamefont {M.}~\bibnamefont
  {Du{\v{s}}ek}}, \bibinfo {author} {\bibfnamefont {N.}~\bibnamefont
  {L{\"u}tkenhaus}},\ and\ \bibinfo {author} {\bibfnamefont {M.}~\bibnamefont
  {Peev}},\ }\bibfield  {title} {\bibinfo {title} {The security of practical
  quantum key distribution},\ }\href@noop {} {\bibfield  {journal} {\bibinfo
  {journal} {Reviews of modern physics}\ }\textbf {\bibinfo {volume} {81}},\
  \bibinfo {pages} {1301} (\bibinfo {year} {2009})}\BibitemShut {NoStop}%
\bibitem [{\citenamefont {Leverrier}\ \emph {et~al.}(2010)\citenamefont
  {Leverrier}, \citenamefont {Grosshans},\ and\ \citenamefont
  {Grangier}}]{leverrier2010finite}%
  \BibitemOpen
  \bibfield  {author} {\bibinfo {author} {\bibfnamefont {A.}~\bibnamefont
  {Leverrier}}, \bibinfo {author} {\bibfnamefont {F.}~\bibnamefont
  {Grosshans}},\ and\ \bibinfo {author} {\bibfnamefont {P.}~\bibnamefont
  {Grangier}},\ }\bibfield  {title} {\bibinfo {title} {Finite-size analysis of
  a continuous-variable quantum key distribution},\ }\href@noop {} {\bibfield
  {journal} {\bibinfo  {journal} {Physical Review A}\ }\textbf {\bibinfo
  {volume} {81}},\ \bibinfo {pages} {062343} (\bibinfo {year}
  {2010})}\BibitemShut {NoStop}%
\bibitem [{\citenamefont {Leverrier}(2015)}]{leverrier2015composable}%
  \BibitemOpen
  \bibfield  {author} {\bibinfo {author} {\bibfnamefont {A.}~\bibnamefont
  {Leverrier}},\ }\bibfield  {title} {\bibinfo {title} {Composable security
  proof for continuous-variable quantum key distribution with coherent
  states},\ }\href@noop {} {\bibfield  {journal} {\bibinfo  {journal} {Physical
  review letters}\ }\textbf {\bibinfo {volume} {114}},\ \bibinfo {pages}
  {070501} (\bibinfo {year} {2015})}\BibitemShut {NoStop}%
\bibitem [{\citenamefont {Wolf}\ \emph {et~al.}(2006)\citenamefont {Wolf},
  \citenamefont {Giedke},\ and\ \citenamefont {Cirac}}]{wolf2006extremality}%
  \BibitemOpen
  \bibfield  {author} {\bibinfo {author} {\bibfnamefont {M.~M.}\ \bibnamefont
  {Wolf}}, \bibinfo {author} {\bibfnamefont {G.}~\bibnamefont {Giedke}},\ and\
  \bibinfo {author} {\bibfnamefont {J.~I.}\ \bibnamefont {Cirac}},\ }\bibfield
  {title} {\bibinfo {title} {Extremality of gaussian quantum states},\
  }\href@noop {} {\bibfield  {journal} {\bibinfo  {journal} {Physical review
  letters}\ }\textbf {\bibinfo {volume} {96}},\ \bibinfo {pages} {080502}
  (\bibinfo {year} {2006})}\BibitemShut {NoStop}%
\bibitem [{\citenamefont {Garc{\'\i}a-Patr{\'o}n}\ and\ \citenamefont
  {Cerf}(2006)}]{garcia2006unconditional}%
  \BibitemOpen
  \bibfield  {author} {\bibinfo {author} {\bibfnamefont {R.}~\bibnamefont
  {Garc{\'\i}a-Patr{\'o}n}}\ and\ \bibinfo {author} {\bibfnamefont {N.~J.}\
  \bibnamefont {Cerf}},\ }\bibfield  {title} {\bibinfo {title} {Unconditional
  optimality of gaussian attacks against continuous-variable quantum key
  distribution},\ }\href@noop {} {\bibfield  {journal} {\bibinfo  {journal}
  {Physical review letters}\ }\textbf {\bibinfo {volume} {97}},\ \bibinfo
  {pages} {190503} (\bibinfo {year} {2006})}\BibitemShut {NoStop}%
\bibitem [{\citenamefont {Weedbrook}\ \emph {et~al.}(2012)\citenamefont
  {Weedbrook}, \citenamefont {Pirandola}, \citenamefont
  {Garc{\'\i}a-Patr{\'o}n}, \citenamefont {Cerf}, \citenamefont {Ralph},
  \citenamefont {Shapiro},\ and\ \citenamefont
  {Lloyd}}]{weedbrook2012gaussian}%
  \BibitemOpen
  \bibfield  {author} {\bibinfo {author} {\bibfnamefont {C.}~\bibnamefont
  {Weedbrook}}, \bibinfo {author} {\bibfnamefont {S.}~\bibnamefont
  {Pirandola}}, \bibinfo {author} {\bibfnamefont {R.}~\bibnamefont
  {Garc{\'\i}a-Patr{\'o}n}}, \bibinfo {author} {\bibfnamefont {N.~J.}\
  \bibnamefont {Cerf}}, \bibinfo {author} {\bibfnamefont {T.~C.}\ \bibnamefont
  {Ralph}}, \bibinfo {author} {\bibfnamefont {J.~H.}\ \bibnamefont {Shapiro}},\
  and\ \bibinfo {author} {\bibfnamefont {S.}~\bibnamefont {Lloyd}},\ }\bibfield
   {title} {\bibinfo {title} {Gaussian quantum information},\ }\href@noop {}
  {\bibfield  {journal} {\bibinfo  {journal} {Reviews of Modern Physics}\
  }\textbf {\bibinfo {volume} {84}},\ \bibinfo {pages} {621} (\bibinfo {year}
  {2012})}\BibitemShut {NoStop}%
\bibitem [{\citenamefont {Pirandola}\ \emph {et~al.}(2017)\citenamefont
  {Pirandola}, \citenamefont {Laurenza}, \citenamefont {Ottaviani},\ and\
  \citenamefont {Banchi}}]{pirandola2017fundamental}%
  \BibitemOpen
  \bibfield  {author} {\bibinfo {author} {\bibfnamefont {S.}~\bibnamefont
  {Pirandola}}, \bibinfo {author} {\bibfnamefont {R.}~\bibnamefont {Laurenza}},
  \bibinfo {author} {\bibfnamefont {C.}~\bibnamefont {Ottaviani}},\ and\
  \bibinfo {author} {\bibfnamefont {L.}~\bibnamefont {Banchi}},\ }\bibfield
  {title} {\bibinfo {title} {Fundamental limits of repeaterless quantum
  communications},\ }\href@noop {} {\bibfield  {journal} {\bibinfo  {journal}
  {Nature communications}\ }\textbf {\bibinfo {volume} {8}},\ \bibinfo {pages}
  {1} (\bibinfo {year} {2017})}\BibitemShut {NoStop}%
\bibitem [{\citenamefont {Pirandola}(2019{\natexlab{a}})}]{pirandola2019end}%
  \BibitemOpen
  \bibfield  {author} {\bibinfo {author} {\bibfnamefont {S.}~\bibnamefont
  {Pirandola}},\ }\bibfield  {title} {\bibinfo {title} {End-to-end capacities
  of a quantum communication network},\ }\href@noop {} {\bibfield  {journal}
  {\bibinfo  {journal} {Communications Physics}\ }\textbf {\bibinfo {volume}
  {2}},\ \bibinfo {pages} {51} (\bibinfo {year}
  {2019}{\natexlab{a}})}\BibitemShut {NoStop}%
\bibitem [{\citenamefont
  {Pirandola}(2019{\natexlab{b}})}]{pirandola2019bounds}%
  \BibitemOpen
  \bibfield  {author} {\bibinfo {author} {\bibfnamefont {S.}~\bibnamefont
  {Pirandola}},\ }\bibfield  {title} {\bibinfo {title} {Bounds for multi-end
  communication over quantum networks},\ }\href@noop {} {\bibfield  {journal}
  {\bibinfo  {journal} {Quantum Science and Technology}\ }\textbf {\bibinfo
  {volume} {4}},\ \bibinfo {pages} {045006} (\bibinfo {year}
  {2019}{\natexlab{b}})}\BibitemShut {NoStop}%
\bibitem [{\citenamefont {Das}\ \emph {et~al.}(2021)\citenamefont {Das},
  \citenamefont {B{\"a}uml}, \citenamefont {Winczewski},\ and\ \citenamefont
  {Horodecki}}]{das2021universal}%
  \BibitemOpen
  \bibfield  {author} {\bibinfo {author} {\bibfnamefont {S.}~\bibnamefont
  {Das}}, \bibinfo {author} {\bibfnamefont {S.}~\bibnamefont {B{\"a}uml}},
  \bibinfo {author} {\bibfnamefont {M.}~\bibnamefont {Winczewski}},\ and\
  \bibinfo {author} {\bibfnamefont {K.}~\bibnamefont {Horodecki}},\ }\bibfield
  {title} {\bibinfo {title} {Universal limitations on quantum key distribution
  over a network},\ }\href@noop {} {\bibfield  {journal} {\bibinfo  {journal}
  {Physical Review X}\ }\textbf {\bibinfo {volume} {11}},\ \bibinfo {pages}
  {041016} (\bibinfo {year} {2021})}\BibitemShut {NoStop}%
\bibitem [{\citenamefont {Mandil}\ \emph {et~al.}(2023)\citenamefont {Mandil},
  \citenamefont {DiAdamo}, \citenamefont {Qi},\ and\ \citenamefont
  {Shabani}}]{mandil2023quantum}%
  \BibitemOpen
  \bibfield  {author} {\bibinfo {author} {\bibfnamefont {R.}~\bibnamefont
  {Mandil}}, \bibinfo {author} {\bibfnamefont {S.}~\bibnamefont {DiAdamo}},
  \bibinfo {author} {\bibfnamefont {B.}~\bibnamefont {Qi}},\ and\ \bibinfo
  {author} {\bibfnamefont {A.}~\bibnamefont {Shabani}},\ }\bibfield  {title}
  {\bibinfo {title} {Quantum key distribution in a packet-switched network},\
  }\href@noop {} {\bibfield  {journal} {\bibinfo  {journal} {npj Quantum
  Information}\ }\textbf {\bibinfo {volume} {9}},\ \bibinfo {pages} {85}
  (\bibinfo {year} {2023})}\BibitemShut {NoStop}%
\bibitem [{\citenamefont {Chen}\ \emph {et~al.}(2023)\citenamefont {Chen},
  \citenamefont {Wang}, \citenamefont {Yu}, \citenamefont {Li},\ and\
  \citenamefont {Guo}}]{chen2023continuous}%
  \BibitemOpen
  \bibfield  {author} {\bibinfo {author} {\bibfnamefont {Z.}~\bibnamefont
  {Chen}}, \bibinfo {author} {\bibfnamefont {X.}~\bibnamefont {Wang}}, \bibinfo
  {author} {\bibfnamefont {S.}~\bibnamefont {Yu}}, \bibinfo {author}
  {\bibfnamefont {Z.}~\bibnamefont {Li}},\ and\ \bibinfo {author}
  {\bibfnamefont {H.}~\bibnamefont {Guo}},\ }\bibfield  {title} {\bibinfo
  {title} {Continuous-mode quantum key distribution with digital signal
  processing},\ }\href@noop {} {\bibfield  {journal} {\bibinfo  {journal} {npj
  Quantum Information}\ }\textbf {\bibinfo {volume} {9}},\ \bibinfo {pages}
  {28} (\bibinfo {year} {2023})}\BibitemShut {NoStop}%
\bibitem [{\citenamefont {Xu}\ \emph {et~al.}(2024{\natexlab{b}})\citenamefont
  {Xu}, \citenamefont {Wang}, \citenamefont {Liao}, \citenamefont {Zhou},
  \citenamefont {Huang},\ and\ \citenamefont {Zeng}}]{xu2024robust}%
  \BibitemOpen
  \bibfield  {author} {\bibinfo {author} {\bibfnamefont {Y.}~\bibnamefont
  {Xu}}, \bibinfo {author} {\bibfnamefont {T.}~\bibnamefont {Wang}}, \bibinfo
  {author} {\bibfnamefont {X.}~\bibnamefont {Liao}}, \bibinfo {author}
  {\bibfnamefont {Y.}~\bibnamefont {Zhou}}, \bibinfo {author} {\bibfnamefont
  {P.}~\bibnamefont {Huang}},\ and\ \bibinfo {author} {\bibfnamefont
  {G.}~\bibnamefont {Zeng}},\ }\bibfield  {title} {\bibinfo {title} {Robust
  continuous-variable quantum key distribution in the finite-size regime},\
  }\href@noop {} {\bibfield  {journal} {\bibinfo  {journal} {Photonics
  Research}\ }\textbf {\bibinfo {volume} {12}},\ \bibinfo {pages} {2549}
  (\bibinfo {year} {2024}{\natexlab{b}})}\BibitemShut {NoStop}%
\end{thebibliography}%
	
\end{document}